\documentclass{aa}
\usepackage{natbib}
\usepackage{graphicx}
\usepackage{txfonts}
\usepackage{color}
\newcommand{\cmtwo}{cm$^{-2}$}  
\newcommand{\cmthree}{cm$^{-3}$}

\newcommand{\kms}{km\,s$^{-1}$}       

\newcommand{\vlsr}{$\upsilon_{\rm LSR}$}        

\newcommand{\tadv}{$\int \! T_{\rm A} d\upsilon$}

\newcommand{\tmbdv}{$\int\!T_{\rm mb}\,d\upsilon$}
\newcommand{\tmb}{$T_{\rm mb}$}

\newcommand{\ecshzsr}{erg cm$^{-2}$ s$^{-1}$ Hz$^{-1}$ sr$^{-1}$}
\newcommand{\um}{$\mu$m}                                 
\newcommand{\molh}{H$_{2}$}                              

\newcommand{\water}{H$_{2}$O}

\newcommand{\lsun}{$L_{\odot}$}                          
\newcommand{\msun}{$M_{\odot}$}
\newcommand{\rsun}{$R_{\odot}$}

\newcommand{\gapprox}{$\stackrel {>}{_{\sim}}$}   
\newcommand{\lapprox}{$\stackrel {<}{_{\sim}}$}
\newcommand{\about}{$\sim$}                       
\newcommand{\powten}[1]{10$^{#1}$}
\newcommand{\ctwo}{[C\,{\sc ii}]\,157\,$\mu$m}    

\newcommand{\oishort}{[O\,{\sc i}]\,63\,$\mu$m}
\newcommand{\oilong}{[O\,{\sc i}]\,145\,$\mu$m}

\newcommand{\ntwohp}{${\rm N_2H^+}$}
\newcommand{\av}{$A_{\rm V}$}                     

\newcommand{\ro}{$\rho \, {\rm Oph}$}
\newcommand{\roc}{$\rho \, {\rm Oph \,\, cloud}$}
\newcommand{\roa}{$\rho \, {\rm Oph \, A}$}

\newcommand{\roac}{$\rho \, {\rm Oph \, A \, core}$}

\newcommand{\amin}{$^{\prime}$}                   
\newcommand{\asec}{$^{\prime \prime}$}
\newcommand{\adeg}{$^{\circ}$}

\newcommand{\atwozero}{$\alpha_{2000}$}
\newcommand{\dtwozero}{$\delta_{2000}$}

\newcommand{\radot}[4]{\mbox{#1$^{\rm h}$#2$^{\rm m}$#3$\stackrel {\rm s}{_{\bf\cdot}}$#4}}  
\newcommand{\decdms}[3]{\mbox{#1$^{\circ}$#2$^{\prime}$#3$^{\prime \prime}$}}
\newcommand{\decdot}[4]{\mbox{#1$^{\circ}$ #2$^{\prime}$ #3$\stackrel {\prime \prime}{_{\bf \cdot}}$#4}}

\newcommand{\amindot}[2]{\mbox{#1$\stackrel {\prime}{_{\bf \cdot}}$#2}}
\newcommand{\asecdot}[2]{\mbox{#1$\stackrel {\prime \prime}{_{\bf \cdot}}$#2}}

\begin{document}

\title{Gas and dust in the star-forming region $\rho$\,Oph\,A
\thanks{Based on observations with APEX, which is a 12\,m diameter submillimetre telescope at 5100\,m altitude on Llano Chajnantor in Chile. The telescope is operated by Onsala Space Observatory, Max-Planck-Institut f\"ur Radioastronomie (MPIfR), and European Southern Observatory (ESO).}$^,$
\thanks{and also based on observations with {\it Herschel} which is an ESA space observatory with science instruments provided by European-led Principal Investigator consortia and with important participation from NASA.}$^,$
\thanks{Figures 2, 11, 12, 16, 17 and 18 are also available in electronic form at the CDS via anonymous ftp to cdsarc.u-strasbg.fr (130.79.128.5) or via http://cdsweb.u-strasbg.fr/cgi-bin/qcat?J/A+A/}
}

\subtitle{The dust opacity exponent $\beta$ and the gas-to-dust mass ratio $g2d$}

\author{
                R. Liseau \inst{1}                                              
        \and
                B. Larsson\inst{2}
        \and
                T. Lunttila\inst{1}
        \and
                M. Olberg\inst{1}
        \and
                G. Rydbeck\inst{1}
        \and
                P. Bergman\inst{1}
        \and
                K. Justtanont\inst{1}
        \and
                G. Olofsson\inst{2}
        \and
                B.L.\,de\,Vries\inst{2, 3}
        }

  \institute{Department of Earth and Space Sciences, Chalmers University of Technology, Onsala Space Observatory, SE-439 92 Onsala, Sweden, 
              \email{\small{rene.liseau@chalmers.se}} 
        \and 
                AlbaNova University Centre, Stockholm University, Department of Astronomy, SE-106 91 Stockholm, Sweden 
        \and 
                Stockholm University Astrobiology Centre, SE-106 91 Stockholm, Sweden
        }

\date{Received ... / Accepted ...}
%
%

\abstract
{}
{We aim at determining the spatial distribution of the gas and dust in star-forming regions and address their relative abundances in quantitative terms. We also examine the dust opacity exponent $\beta$ for spatial and/or temporal variations.}
{Using mapping observations of the very dense \roac, we examined standard 1D and non-standard 3D methods to analyse data of far-infrared and submillimeter (submm) continuum radiation. The resulting dust surface density distribution can be compared to that of the gas. The latter was derived from the analysis of accompanying molecular line emission, observed with {\it Herschel} from space and with APEX from the ground. As a gas tracer we used \ntwohp,  which is believed to be much less sensitive to freeze-out than CO and its isotopologues. Radiative transfer modelling of the \ntwohp\,(J=3-2) and (J=6-5) lines with their hyperfine structure explicitly taken into account provides solutions for the spatial distribution of the column density $N$(\molh), hence the surface density distribution of the gas.}
{The gas-to-dust mass ratio is  varying across the map, with very low values in the central regions around the core SM\,1. The global average, $=88$, is not far from the canonical value of 100, however. In \roa, the exponent $\beta$ of the power-law description for the dust opacity exhibits a clear dependence on time, with high values of 2 for the envelope-dominated emission in starless Class\,--1 sources to low values close to 0 for the disk-dominated emission in Class\,III objects. $\beta$ assumes intermediate values for evolutionary classes in between.}
{Since $\beta$ is primarily controlled by grain size, grain growth mostly occurs in circumstellar disks. The spatial segregation of gas and dust, seen in projection toward the core centre, probably implies that, like C$^{18}$O, also \ntwohp\ is frozen onto the grains.}

\maketitle
\section{Introduction}

\begin{figure}
  \resizebox{\hsize}{!}{
    \rotatebox{0}{\includegraphics{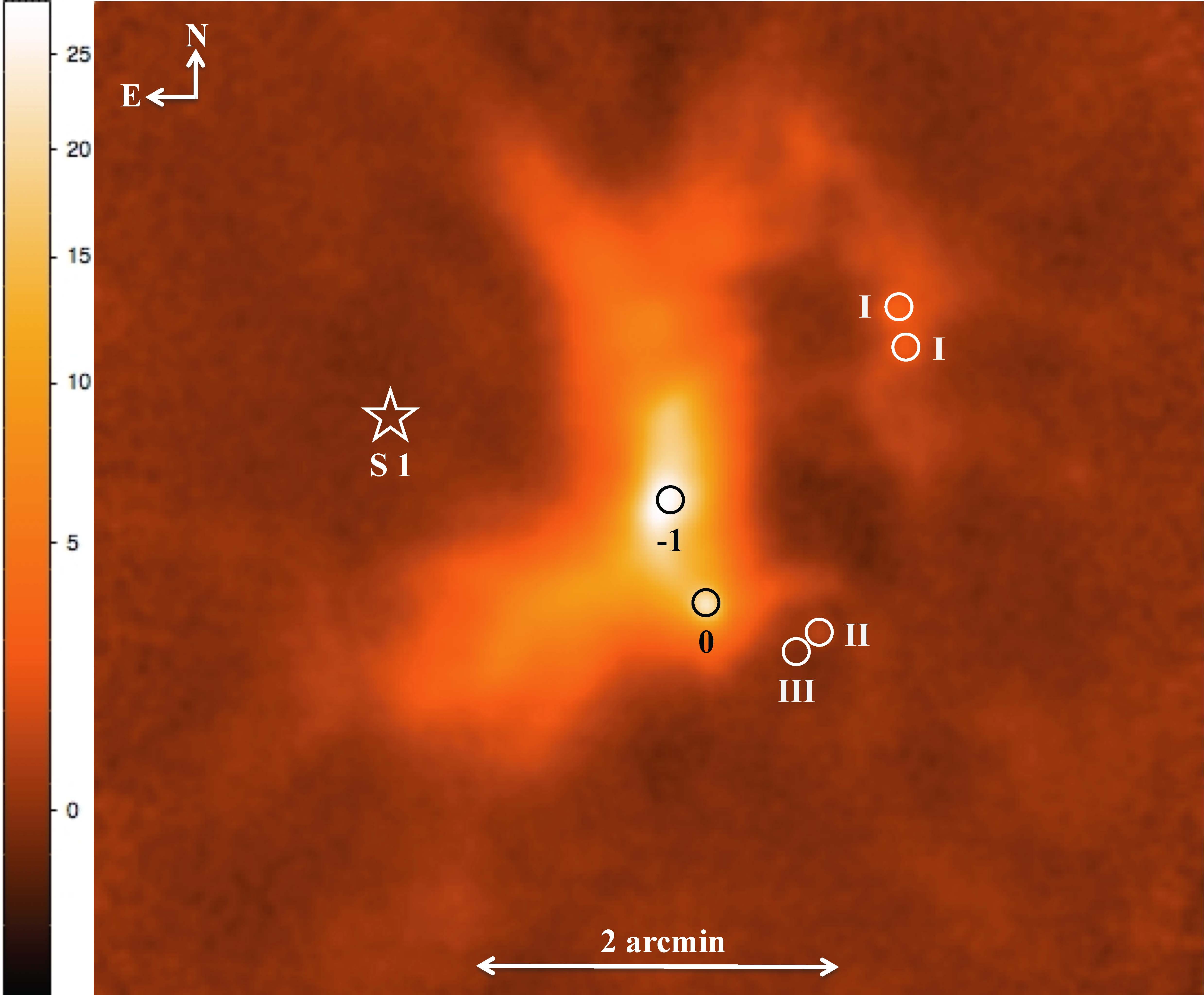}}     
                        }
  \caption{The core \roa\ in the continuum at 350\,\um, where the different evolutionary phases are identified. The position of the dominating radiative source S\,1 is shown by the star symbol. The colour scale is in units of Jy\,beam$^{-1}$, with the circles corresponding to the beam size. The angular scale is shown at the bottom of the figure and the orientation in the sky in the upper left corner. The APEX maps of this paper have their origin at R.A. = \radot{16}{26}{27}{90}, Dec = \decdot{$-24$}{23}{57}{00} (J2000.0). 
  }
    \label{classes}
\end{figure}

\begin{figure}
  \resizebox{\hsize}{!}{
    \rotatebox{270}{\includegraphics{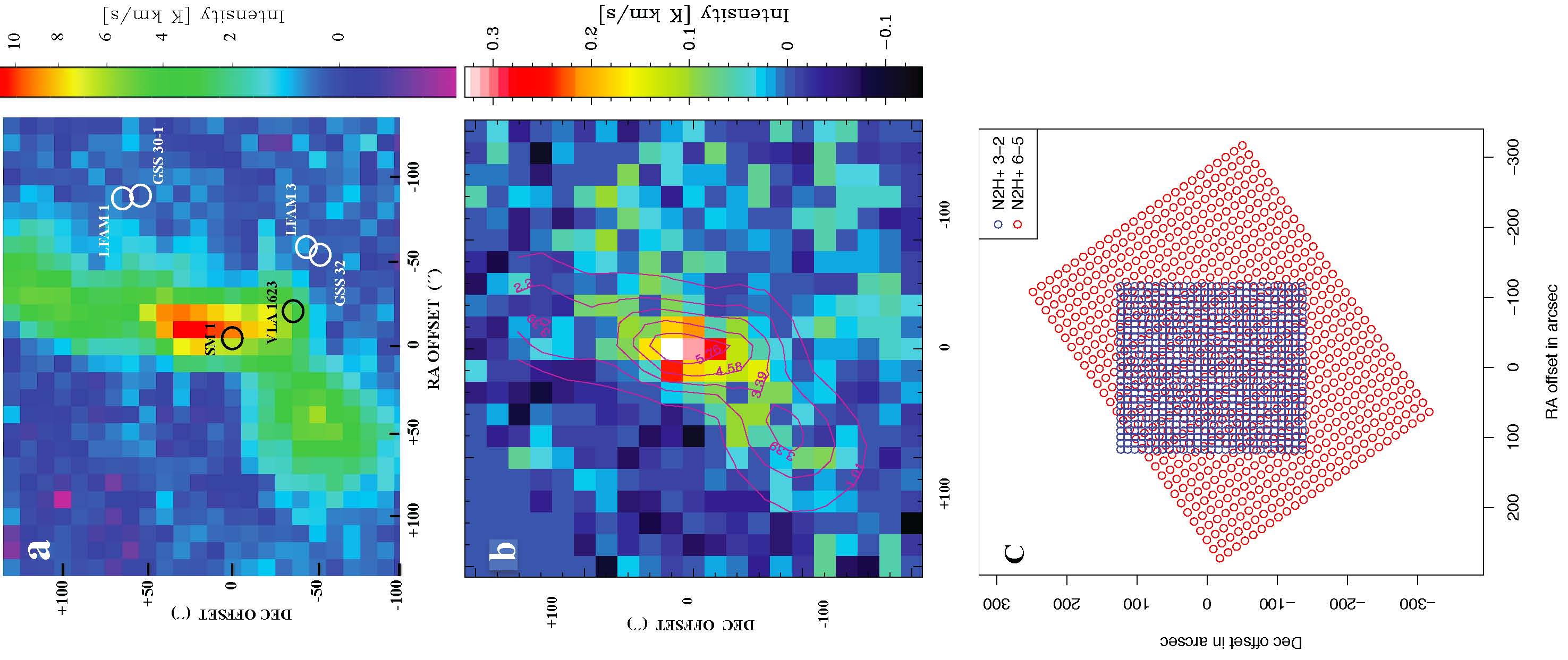}}     
                        }
  \caption{{\bf (a)} Integrated intensity, \tadv, of N$_2$H$^+$\,(3-2) at 22\asec\ resolution with APEX and with a sampling rate of 10\asec/pixel is shown in the upper panel. The positions of the SED-class sources of Fig.\,\ref{classes} are shown with their proper designations.  {\bf (b)} The {\it Herschel} map of the \roac\ in the N$_2$H$^+$\,(6-5) line that is rotated to align with the APEX data. The (6-5) map has a resolution of 38\asec\  and is sampled at a rate of 16\asec/pixel. Superposed, and shown as contours, is the map in the (3-2) line at degraded resolution. {\bf (c)} The orientation of the maps in the sky.
   }
  \label{N2H+3_2}
\end{figure}


It is widely accepted that the gas-to-dust mass ratio ($g2d$) in our Galaxy has a value of about 100. Of course, this is an average value and it does, in particular, pertain to the {\it \textup{diffuse}} interstellar medium (ISM). However, this value is also commonly adopted for dense cloud environments, such as star-forming regions, and the methods used to determine that parameter are described for instance by \citet{kenyon1998}, \citet{draine2003}, and  \citet{vuong2003}. Based on observations of CO isotopologues and stellar extinction values in the nearby star-forming regions in Taurus and Ophiuchus, \citet{frerking1982} established the still widely used molecular gas-dust relations, that is, $N(^{X}$C$^{Y}$O)$\,\propto\,$\av, \av\,\lapprox\,20\,mag, and $^{12}$C$^{16}$O/\molh\,$\sim\,8.5\times 10^{-5}$. For higher \av, saturation makes these relations less useful.

\citet{young1991} discussed the high average $g2d$ of \about\,600 obtained for molecular clouds, where gas masses refer to H\,I + \molh. The latter was derived from CO observations and the dust masses were determined from IRAS observations. Furthermore, based on data for H\,I and H\,II regions, \citet{spitzer1978} quoted the rather wide range of 20 to 700 for this parameter. A value of about 450 was found by \citet{liseau1995} for L\,1688 using CS maps and 60\,\um\ IRAS data. A high $g2d$ would also be consistent with the low oxygen abundance in \roa\ found by \citet{liseau2009}.  Figure\,11 of \citet{brinchmann2013}, which is based on a large number of galaxies, supports this view on larger, extragalactic scales. 

One of the reasons for these high $g2d$ values could be that large portions of the dust are at temperatures considerably lower than 30\,K and, hence, remained undetected by IRAS \citep{young1991}. However, at temperatures below about 25\,K, the abundance of the dominant gas tracer, CO and its isotopologues, could also
become highly diminished by freeze-out  \citep[e.g.,][and references therein]{lippok2013}.  One might therefore select spectral lines of a molecule that is seen in regions where CO is thought to be depleted. One such molecule is  diazenylium, \ntwohp\ \citep[e.g.,][]{caselli1999}, which is formed from N$_2$ (which remains in the gas phase to somewhat lower temperatures than CO) and is removed from the gas phase mainly by CO \citep{bergin2002,oberg2005}. However, as pointed out by \citet{bergin2002}, even \ntwohp\ (and N$_2$) may become depleted and thus leave very few molecular probes \citep[perhaps only H$_2$D$^+$,][]{brunken2014} to study the centres of the densest cores.

The rotational transitions (J=3-2) and (J=6-5) of \ntwohp, at 280\,GHz and 560\,GHz and with upper level energies of 27\,K and 94\,K, respectively, could fulfil these requirements. In particular, the (1-0) line at 93\,GHz, having an upper level energy similar to that of CO, has been widely used for dense interstellar clouds, including the \roa\ region  \citep[e.g.,][]{difrancesco2004}. However, $E_1/k$ is only 4.5\,K and at considerably higher temperatures, the population of the $J=0$ and $J=1$ levels changes only slowly and does as such not contribute much to the understanding of the overall excitation of the molecule.

\citet{friesen2014} have used the Atacama Large Millimeter/submillimeter Array (ALMA) to observe \roa\ in the \ntwohp\,(4-3) lines toward a small region ($< 20^{\prime\prime} \times 20$\asec), but the paper focused on the continuum data. There are no line profiles nor  line maps shown for these data, which, according to \citet{friesen2014},  were either self-absorbed or resolved out by the interferometer. The authors also noted that line and continuum sources did not coincide, neither for SM\,1 nor for SM\,1N, a point of potential importance (see below, Sect.\,4.4). The primary beam of these observations is smaller than the single-dish beams of our observations.

As revealed by pure-rotational \molh\ line emission, relatively high temperatures, \gapprox\,\powten{3}\,K, are present  in the interface regions toward the dominating radiation source S\,1 (Fig.\,\ref{classes}). In addition, \citet{liseau1999} found extended \oishort, \oilong\ and \ctwo\ emission, which is indicative of temperatures in excess of 100 K. These phenomena probably refer to the outer layers of the cloud. The \ntwohp\,(6-5) line could be used to probe that temperature regime in a coherent  map.

\begin{figure*}
\centering
    \rotatebox{270}{\includegraphics[width=23.0cm]{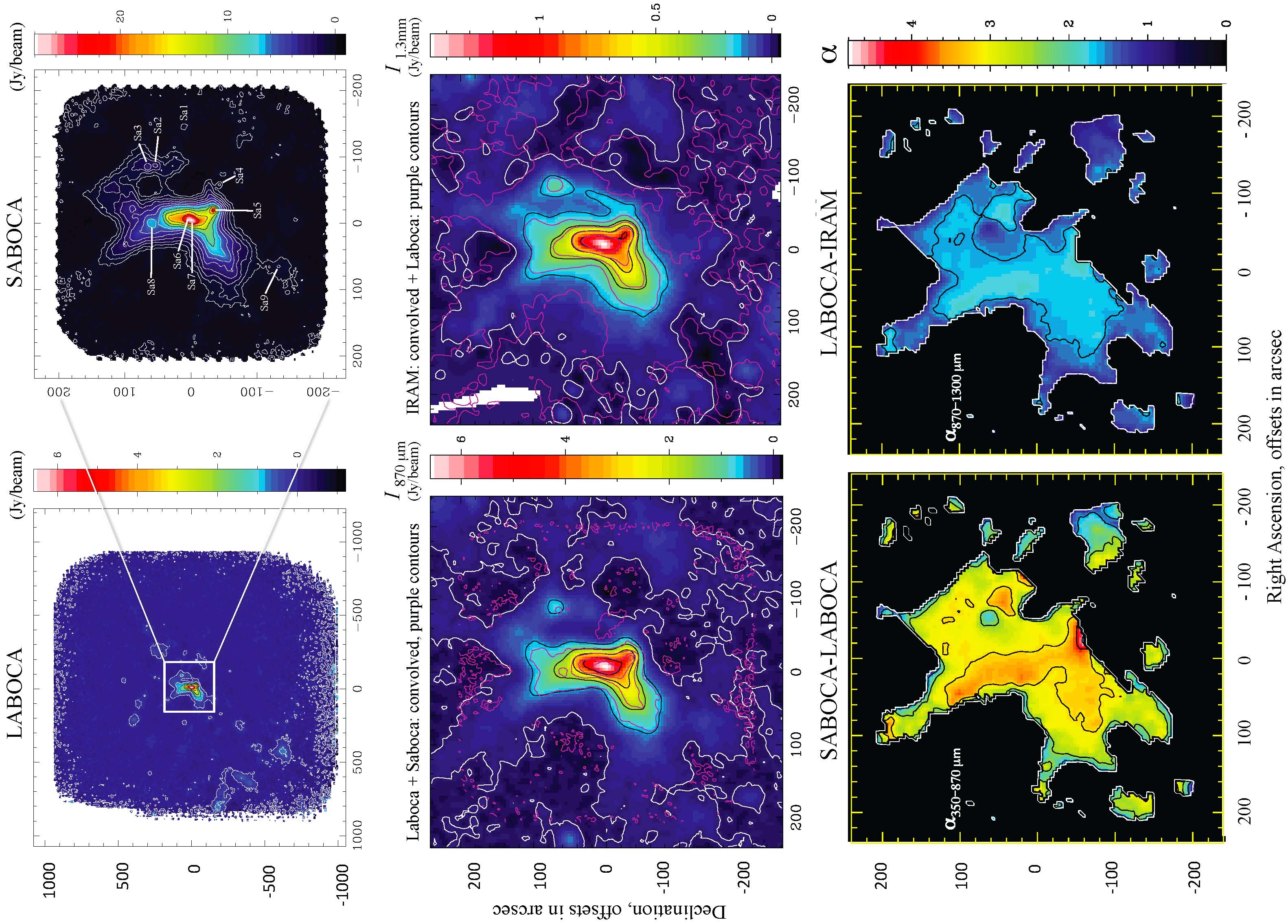}}                        
  \caption{{\bf Top:} \roa\ observed at 870\,\um\ with LABOCA (left) and at 350\,\um\ with SABOCA (right). Offsets are in seconds of arc and units are in Jy/beam. The origin, (0,\,0), is the same as in Fig.\,\ref{classes} and the numbers refer to Table\,\ref{saboca_pos}. {\bf Middle:} The SABOCA data (left, purple contours), degraded to the \asecdot{19}{5} resolution of the LABOCA map (in colour).  The colours show the IRAM map (right) of \citet{motte1998}, convolved to \asecdot{19}{5} resolution, and shifted to the position of the LABOCA origin. The LABOCA data are shown as purple contours.  {\bf Bottom:} The observed spectral index $\alpha = \Delta \log I_\nu/ \Delta \log \nu$ for the wavebands 350\,\um--870\,\um\ (left) and 870\,\um--1.3\,mm (right).}
  \label{bocas}
\end{figure*}

Already early on, the actively star-forming \roc s attracted wide attention, largely because the nearest stellar  cluster is forming there. This is the dominant star formation mode, in contrast to more isolated star formation as in  the Taurus-Auriga clouds, for example, that is  contributing only little to the stellar initial mass function (IMF) of the Galaxy. Not surprisingly therefore, a huge body of literature exists, and we refer to the review by \citet{wilking2008} and also to a few recent specific papers on the distribution of molecular gas and icy grains \citep{shuping2000,pontoppidan2006,andre2007,white2015}. For these clouds in \ro, the visual extinction can locally reach extremely high values, $(\max{A_{\rm V}} > 10^2$\,mag).

The nearby 3\amin\ (0.1\,pc) \roac\  \citep{loren1990} is unique in the sense that {\it \textup{all}} phases of the earliest phases of stellar evolution are present in a very compact region (see Figs.\,\ref{classes} and \ref{N2H+3_2}) - from starless clumps, Class\,$-1$ (SM\,1), via Class\,0 (VLA\,1623), Class\,I (GSS\,30-1) and Class\,II (LFAM\,3) to Class \,III (GSS\,32) \citep[see, e.g.,][]{comeron1993,bontemps2001,white2015}. Potentially, this fact may make it possible to follow the time evolution of the dust in various locations in the core, including grain growth and changes in composition, such as ice coatings onto initially ``bare'' grains. Such a scenario could have considerable effect on theories of planetary formation (e.g., the location of the {\it \textup{snow line}} in the circumstellar disks) and help to relax tight timescale requirements, particularly in the outer, less dense regions of the disks. The status of the current understanding of the dust evolution in disks has recently been summarized by  \citet{testi2014}.

In this paper, we focus on the gas and dust contents of the nearby star-forming \roa\ core, observed in the far-infrared (FIR) and sub-millimeter spectral regions (submm). For the distance we adopt the accurately determined value with the Very Long Baseline Array (VLBA) by \citet[][viz. $D=120.0^{+4.5}_{-4.2}$\,pc]{loinard2008}. The organization is as follows: in Sect.\,2, our mapping observations of freeze-out insensitive molecules are described. These observations have been obtained from space with {\it Herschel} \citep{pilbratt2010} and from the ground with APEX at 5100\,m altitude. There, maps in the submm continuum have also been obtained. This section also describes the reduction of these data, and the results are presented in Sect.\,3. These are discussed in Sect.\,4, illustrating specifically the aspects of spectral line overlap and also of geometry in the radiative transfer problems. We compare the results obtained with tools widely in use with those arrived at with our own developments. The conclusions follow in Sect.\,5. 

\section{Observations and data reduction}

\begin{table*}
    \caption{Log of observations}
    \label{obs_log}
    \begin{tabular}{llclllrcr}      
\hline\hline    
\noalign{\smallskip}             \noalign{\smallskip}   
Instrument              & Prog/Obs ID           &  Mode                         & Frequency       & Beam                                  & Observing Date    & $t_{\rm int}$                 & Map Dimension &       $P.A.$  \\
                                & ESO/{\it Herschel}    &                                       &  (GHz)          & HPBW                                  &  yy--mm--dd           & (sec)                           & R.A.$ \times$ Dec. &                  \\
\noalign{\smallskip}    
\hline                          
\noalign{\smallskip}    
APEX-2$^{a}$        & 090.F-9319(A) & N$_2$H$^+\,(3-2)$     & 279.515067      & 22\asec                                       &  12--09--22 to 25       &   65160                       & \amindot{4}{26} $\times$ \amindot{4}{32}        & 0\adeg        \\
HIFI$^{b}$              & 1342204010$^{c}$& N$_2$H$^+\,(6-5)$         & 558.957500    & 38\asec                                       &  10-09-01               &   \phantom{1}6552     & \amindot{5}{0} $\times$ \amindot{2}{0}         &35\adeg        \\
                                & 1342251245$^{d}$&                                     &                       &                                               &  12-09-21               &    \phantom{1}3608& \amindot{5}{8} $\times$ \amindot{1}{0}         &35\adeg        \\                              
                                & 1342251246$^{e}$&                                     &                       &                                               &  12-09-21               &   \phantom{1}5712& \amindot{5}{0} $\times$ \amindot{2}{0}          & 125\adeg  \\          
                                & 1342251247$^{f}$&                                     &                       &                                               &  12-09-21               &   \phantom{1}5712     &  \amindot{5}{0} $\times$ \amindot{2}{0}         & 125\adeg \\           
                                & 1342251429$^{g}$&                                     &                       &                                               &  12-09-26               &   \phantom{1}3608     & \amindot{5}{8} $\times$ \amindot{1}{0}         & 35\adeg         \\
LABOCA$^{a}$            & 090.F-9304(A) & Continuum                     & 345             & \asecdot{19}{5}                       &  12--08--17           &   \phantom{1}9298       & \amindot{38}{3} $\times$ \amindot{35}{5}      & 0\adeg  \\
SABOCA$^{a}$    & 090.F-9304(B) & Continuum                     & 852             & \phantom{1}\asecdot{7}{5}     &  12--08--22           &   20419                         & \amindot{7}{75} $\times$ \amindot{7}{85}      & 0\adeg  \\
\noalign{\smallskip}    
\hline
    \end{tabular}
    \begin{list}{}{}
    \item[$^{a}$] Map offsets refer to \atwozero\,= \radot{16}{26}{27}{90}, \dtwozero\,= \decdot{$-24$}{23}{57}{0}.
    \item[$^{b}$] The HIFI data refer to OD\,475 on 10-09-01 and OD\,1221 on 12-09-16  and OD\,1231 on 12-09-26, with a total integration time of  25192\,s.  
    \item[$^{c}$] Observed centre coordinates are  \atwozero\,= \radot{16}{26}{26}{38}, \dtwozero\,= \decdms{$-24$}{24}{31}; [$^{d}$]  \radot{16}{26}{13}{79}, \decdms{$-24$}{22}{31}; [$^{e}$] \radot{16}{26}{21}{30}, \decdms{$-24$}{26}{15};  [$^{f}$]  \radot{16}{26}{31}{73}, \decdms{$-24$}{22}{47};  [$^{g}$]  \radot{16}{26}{38}{97}, \decdms{$-24$}{26}{32}.   
    \end{list}
\end{table*}

\subsection{SHeFI-APEX-2 and {\it Herschel}-HIFI}

\subsubsection{APEX-2}

The Atacama Pathfinder Experiment (APEX) is a 12\,m single-dish telescope situated at 5100\,m altitude in Chile. The pointing accuracy of the telescope has been determined at 2\asec (rms). We used the Swedish heterodyne facility instrument APEX-2 \citep{vassilev2008} for the raster-mapping observations of \roa\ in the $J=3-2$ rotational transition of the N$_2$H$^+$ molecule. The central rest frequency is at 279515.06711\,MHz and the average zenith optical depth was 0.04. At 279.5\,GHz, the beam size is FWHM = 22\asec, the main beam efficiency is $\eta_{\rm mb}=0.73$, and the LSB receiver temperature was 125\,K. We used the Fast Fourier Transform Spectrometer with 73.6\,kHz wide channels, resulting in a velocity resolution of 0.082\,\kms\ per channel and a spectral coverage corresponding to 200\,\kms. The data reduction involved the fitting and removal of the baselines and was made with the locally available software package {\texttt{xs}\footnote{\texttt{\small{ftp://yggdrasil.oso.chalmers.se/pub/xs/}}}.

\subsubsection{HIFI}

The ($J$=6-5) line of \ntwohp\ was observed at the same time as our {\it Herschel} \water\ maps (2012, GT2\_rliseau\_1). The observational details regarding the partial map data of 2010 (Table\,\ref{obs_log}) that were obtained within the KPGT\_evandish\_1 programme were already described in detail by \citet{bjerkeli2012}.  In their Fig.\,1, offsets are relative to the (0, 0)-position at  \atwozero\,=\,\radot{16}{26}{26}{4}, \dtwozero\,=\,\decdms{$-24$}{24}{31}. The more extended mapping observations  in 2012 were made on-the-fly, with a reference position 14\amin\ away. The data reduction was made within HIPE\,10\footnote{{\it Herschel} Interactive Processing Environment, see \\ \texttt{\small{herschel.esac.esa.int/hipe/}}}. For the final composite map, the zero-offset is at \atwozero\,=\,\radot{16}{26}{27}{78}, \dtwozero\,=\,\decdot{$-24$}{23}{50}{3}, which is slightly offset from the APEX origin by (\asecdot{$-1$}{6},\,\asecdot{$-6$}{7}) and which is smaller than the beam size at 559\,GHz, that is, 38\asec. At this frequency in HIFI-band\,1, the main beam efficiency\footnote{Mueller et al.\,(2014): Release Note \#1, HIFI-ICC-RP-2014-001, v1.1} is $\eta_{\rm mb}=0.63$. 

\subsection{LABOCA and SABOCA}

The continuum observations were carried out with two bolometer multi-pixel cameras, viz. the Large Apex Bolometer Camera LABOCA \citep{siringo2009} and the Submillimetre Array, that is, SABOCA \citep{siringo2010}. The effective operating frequencies correspond to 870\,\um\ and 350\,\um, with about 150 and 50\,\um\ wide filters, respectively\footnote{{\texttt{http://www.apex-telescope.org/instruments/}}}. The resolutions are \asecdot{19}{5} and \asecdot{7}{5}, respectively. At the two wavelengths, the maps of \roa\ span almost 40\amin\,$\times 40$\amin\ and 8\amin\,$\times 8$\amin, respectively (Fig.\,\ref{bocas}), and in both maps the centre coordinates are at R.A. = \radot{16}{26}{27}{90} and Dec = \decdot{$-24$}{23}{57}{0} for the epoch of J2000.0. 

The \amindot{11}{4}  circular field of view (FOV) of LABOCA is determined by the 295 bolometers, which are arranged with  36\asec\ spacings in between them. The number of loosely spaced bolometers  of SABOCA is 39 and its FOV is \amindot{1}{5}. The FOV is under-sampled for both cameras and is filled by repetitive continuous scanning of the mapped region. For both LABOCA and SABOCA, the data reductions were made with CRUSH\footnote{ {\small{\texttt{http://www.submm.caltech.edu/sharc/crush/download.htm}}}}. CRUSH uses the actual atmospheric opacities at the time of observation. For the two data sets, average zenith optical depths were 0.42 at 870\,\um\ and 1.07 at 350\,\um, with a range of 0.55 to 1.20 for the latter.

\begin{figure*}
  \resizebox{\hsize}{!}{
    \rotatebox{00}{\includegraphics{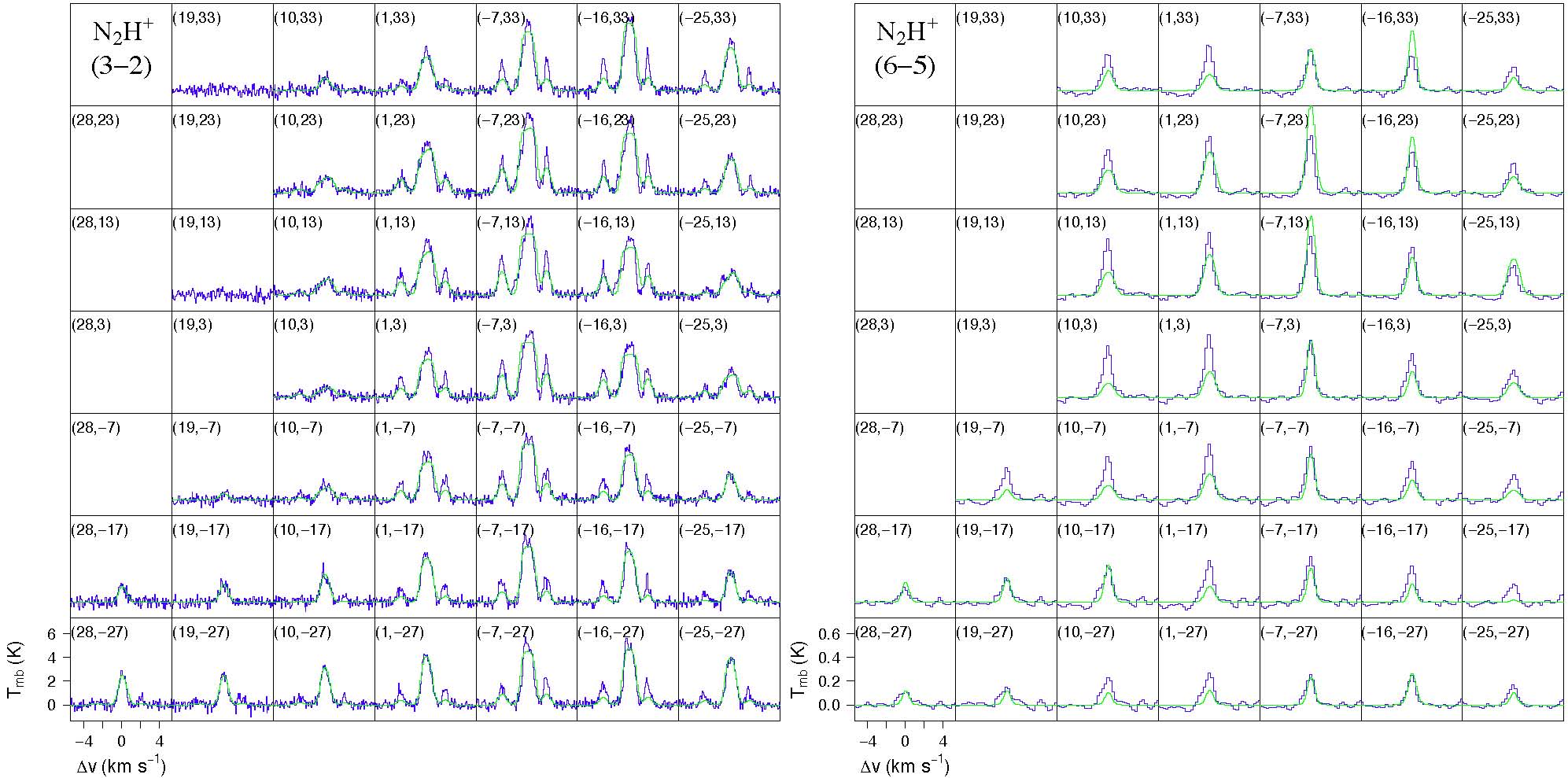}}    
                        }
  \caption{Centre of the three sub-regions of the mapped area showing the best-fit models (green) for the observed (blue) \ntwohp\,(3-2) and (6-5) line profiles (left and right, respectively). The radiative transfer with ALI  takes the overlapping transitions of the 38 hyperfine components into account, with the frequencies of \citet{pagani2009}. In the small frames, the numbers in the upper left corners are equatorial offsets in arcsec from the map origin (0,\,0). In the lower left frames, the scales are given for \tmb\,(K) and $\Delta$v $= \upsilon - \upsilon_{\rm LSR}$ (\vlsr=+3.5\,\kms), respectively (Figs.\,\ref{aliresults} and \ref{line_overlap}).}
  \label{hfs_width}
\end{figure*}

\begin{figure*}[ht]
  \resizebox{\hsize}{!}{
    \rotatebox{00}{\includegraphics{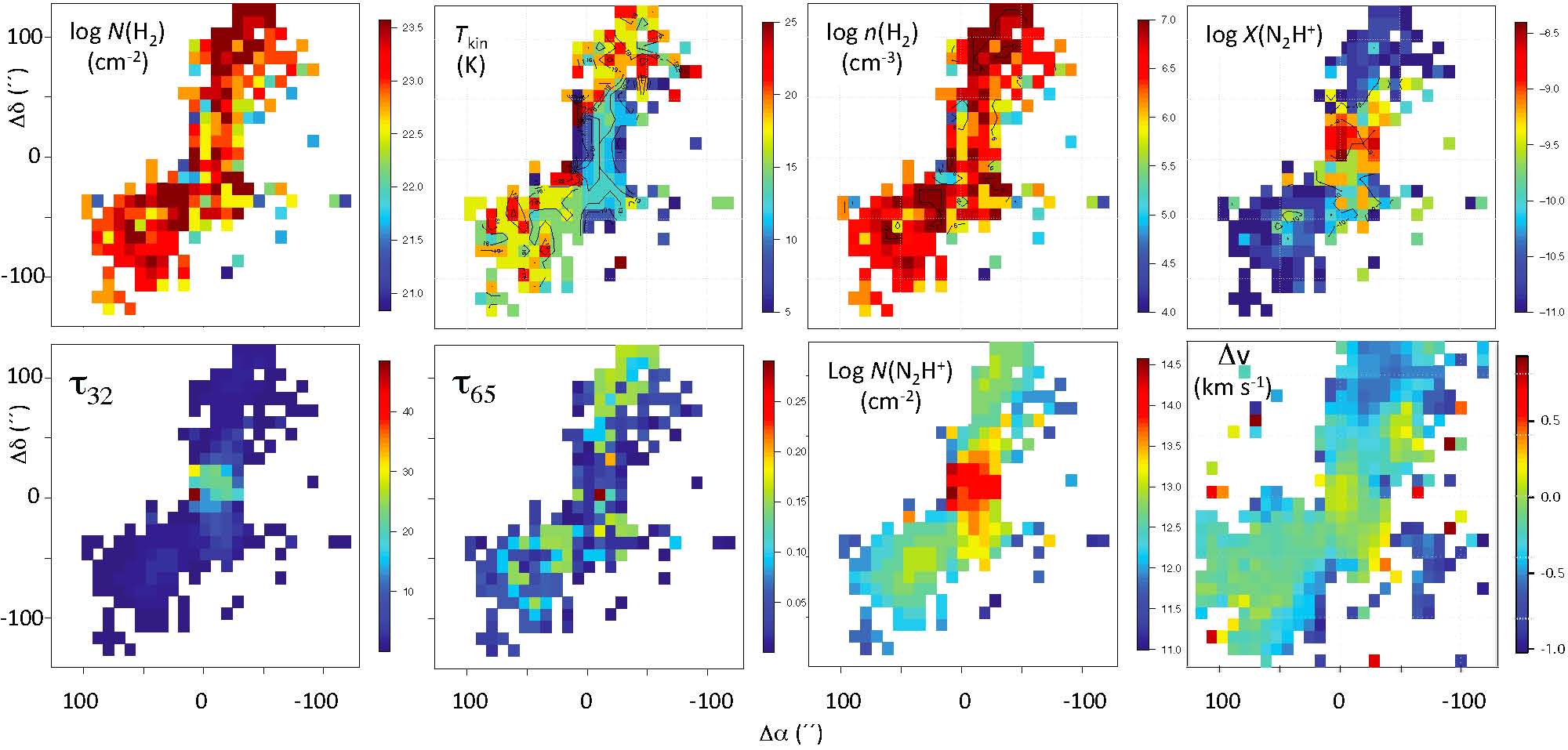}}
                 }
  \caption{Best-fit parameters from the ALI modelling of the observed \ntwohp\ maps, yielding results as shown in Fig.\,\ref{hfs_width}. Modelled data use an intensity threshold of  $3\sigma$ of the rms noise for both lines. {\bf Upper row:} From left to right, the spatial distributions are shown for the gas column density, $N$(\molh), the kinetic gas temperature, $T_{\rm kin}$, the gas volume density, $n$(\molh) and the relative molecular abundance, $X$(\ntwohp)=$n$(\ntwohp)/$n$(\molh).  {\bf Lower row: }From left to right, the spatial distributions are shown for the line centre optical depth of the (3-2) and (6-5) lines, the \ntwohp\ column density, $N$(\ntwohp), and the residuals of the radial velocity fits $\Delta$v in \kms. Spatial scales are in arcsec and the colour coding is given by the vertical bars next to each frame. 
 }
  \label{aliresults}
\end{figure*}

\section{Results}

\subsection{Spectral line data}

The maps of main-beam corrected integrated intensity, \tmbdv, of \ntwohp\,(6-5) and (3-2) are shown in Fig.\,\ref{N2H+3_2}. The mapping of the (6-5) line with HIFI was made at an angle with respect to the equatorial coordinate system, so that the map had to be rotated to be properly aligned with the APEX data (Fig.\,\ref{N2H+3_2}c). In Figs.\,\ref{N2H+3_2}a and b, the shown pixel sizes correspond to the respective samplings, and in b, the superposed contours show the (3-2) data after convolution to the (6-5) resolution (38\asec).  

\subsection{Continuum data}

The reduced data are shown in Fig.\,\ref{bocas} for both LABOCA and SABOCA, where intensity units are Jy\,beam$^{-1}$. These intensity scales assume Gaussian beams given in Table\,\ref{obs_log}, and  in the figure,  the contour levels for LABOCA are (1) -1.21, (2) 0.13,  (3)  0.59,  (4) 1.20, (5)  2.02, (6)  3.13, (7)  4.61, (8)  6.60 Jy\,beam$^{-1}$. For SABOCA, the corresponding contours are (1) -1.01, (2)  0.31, (3)  0.74, (4)  1.31, (5)  2.06, (6)  3.06, (7)  4.38,  (8)  6.12,  (9)  8.44, (10) 11.50, (11) 15.55, (12) 20.92, (13) 28.03 Jy\,beam$^{-1}$. The value of 1\,Jy\,beam$^{-1}$ corresponds to an intensity of about 100\,MJy\,sr$^{-1}$ for LABOCA and 600\,MJy\,sr$^{-1}$ for SABOCA, and with $1\sigma$ of the rms noise corresponding to 5 and 30\,MJy\,sr$^{-1}$, respectively. 

At the higher resolution of the SABOCA observations, individual sources can be identified, and their equatorial coordinates are presented in Table\,\ref{saboca_pos}. The rms pointing accuracy of the APEX telescope is about 2\asec.
The observed field at 350\,\um\ is much smaller than that at 870\,\um\ (Fig.\,\ref{bocas}). For quantitative applications, the SABOCA data have to be degraded to the appropriate LABOCA values, that is,  \asecdot{19}{5}, and re-sampled onto a 4\asec\ grid. This is illustrated in the middle panels of Fig.\,\ref{bocas}, where SABOCA contours are shown superposed on the LABOCA intensities. In addition, the IRAM 1.3\,mm data of  \citet{motte1998} are shown, also convolved to the \asecdot{19}{5} resolution of LABOCA and sampled onto the 4\asec\ regular grid. For the comparison with the APEX results, the map of \roa\ was shifted to the same centre coordinates.

\begin{table}
\caption{SABOCA objects at \asecdot{7}{5} resolution}             
\label{saboca_pos}      
\begin{tabular}{clll}      
\hline\hline    
\noalign{\smallskip}             
ID      &  R.A. (J2000)         & Dec. (J2000)                  &       Compactness/Source      \\
        &  hh mm ss.sss & \phantom{1}\adeg\  \phantom{1}\amin\  \phantom{1}\asec       &               \\
\noalign{\smallskip}    
\hline                        
\noalign{\smallskip}                            
Sa1             & 16 26 17.30 & $-24$ 23 46.5                   & diffuse, ISO-Oph 21\\
Sa2             & 16    26 21.53 & $-24$ 23 04.5                        & point, GSS 30-1 \\
Sa3             & 16 26 21.75 & $-24$ 22 51.0                   & point, LFAM\,1 \\
Sa4             & 16 26 23.67 & $-24$ 24 39.8                   & point, LFAM\,3 \\
Sa5             & 16 26 26.42 & $-24$ 24 30.4                   & point, VLA 1623 \\
Sa6             & 16 26 27.49 & $-24$ 23 56.2                   & diffuse, SM 1N\\
Sa7             & 16 26 27.52 & $-24$ 23 54.8                   & diffuse, SM 1\\
Sa8             & 16 26 27.74 & $-24$ 22 55.5                   & diffuse \\
Sa9             & 16 26 32.68 & $-24$ 26 09.8                   & diffuse \\      
\hline                                   
\end{tabular}
\end{table}

 \begin{figure}
  \resizebox{\hsize}{!}{
    \rotatebox{00}{\includegraphics{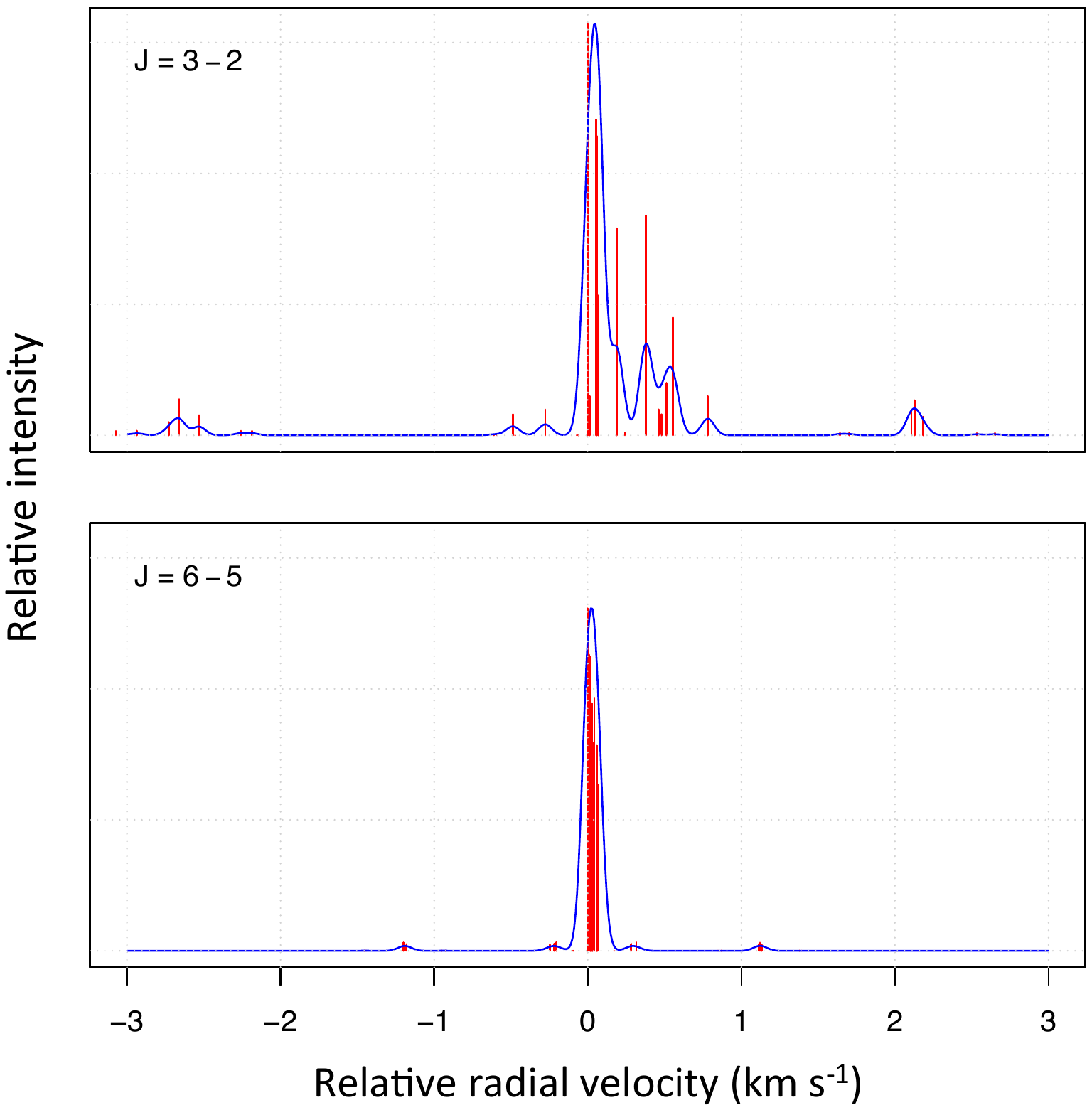}}
                        }
  \caption{Hyperfine line strengths (red vertical bars) for the $J=3-2$ (upper panel) and $J=6-5$ (lower panel) transitions. The blue curves are theoretical profiles for an optical depth of 0.1 and a line width of 0.1\,\kms, scaled to the same maximum amplitude as the LTE line strengths.}
  \label{line_overlap}
\end{figure}

\section{Discussion}

We start our discussion with determining the physical parameters for the gas, in particular $T_{\rm kin}$ and $N$(\molh), using spectral line maps of (presumably) optically thin emission, and C$^{18}$O has often been used to guarantee this. However, as shown by \citet{liseau2010} on the basis of C$^{18}$O\,(3-2) and  $^{13}$C$^{18}$O\,(3-2) observations of \roa, optical depths do exceed unity even in the rare isotopologues of CO. In those cases, the observed radiation does not carry information about the entire core but is restricted to layers closer to the surface, leading to erroneous estimates of the overall temperature and density distribution. For instance, temperatures based on such observations are around the (for a dense core) surprisingly high value of \about\,25\,K. At temperatures below this, which are commonly encountered in dense cores, the CO gas will be depleted because of freeze-out of the molecules onto the surfaces of the accompanying grains. In contrast, \ntwohp\ molecules remain in the gas phase down to much lower temperatures \citep[][and references therein]{lippok2013}. Except for a few exceptions, only the (1-0) line and its hyperfine components have been observed and the analysis generally assumed thermodynamic equilibrium (LTE). Here, we make multi-transition analyses and treat the radiative transfer in non-LTE. The implications for our observations are described in the next section.

\subsection{Gas temperature, opacity, and mass}

For large numerical grids, the publicly available computer code \texttt{radex} \citep{vdtak2007} is very suitable. In its present form, the code is unable to properly handle the transfer in optically thick lines with extremely closely spaced transitions,
however (the ``line overlap'' problem, Fig.\,\ref{line_overlap}). In addition, by construction, the code will not return the proper shapes of the lines. To surmount these difficulties, we used for the analysis an accelerated lambda iteration (ALI) code that takes line-overlap explicitly into account and also correctly computes the line profiles with which the observations can be compared directly. 

The upper level energy of the \ntwohp\,(3-2) transition corresponds to nearly 28\,K, close to previous estimates of the kinetic gas temperature. The observed (3-2) lines exhibit the remarkable feature that the satellite lines reveal strengths that are almost similar to that of the main line (Fig.\,\ref{hfs_width}), that
is, much stronger than expected for optically thin radiation in LTE (Fig.\,\ref{line_overlap}).  In fact, this strongly suggests optically very thick emission in the overlapping components of the central, spectrally non-resolved, main line. This line anomaly is in contrast to the observed (6-5) line profile, which appears nearly Gaussian (however, the spectral resolution is lower).  The $J=6$ level is about 94\,K above ground.

The Onsala ALI code (P.\,Bergman) is based on the work by \citet{rybicki1991} and  \citet{rybicki1992} and the bench-marking has been reported by \citet{maercker2008}. It  takes line overlap explicitly into account and, in the following, we use this program to model the multitude of hfs lines of \ntwohp. The geometry is spherically symmetric, which needs to be kept in mind when comparing model line profiles with observed ones. Dust continua are included, but switched out in the simulated on-off observations presented below (zero baseline).

The radiative  transfer in the hfs transitions with line overlap was computed using the energy levels that we derived from the frequencies  presented by \citet{pagani2009}. The collisional de-excitation rate coefficients were adopted from \citet{lique2015}, who have calculated the rates for collisions with \molh\ for the temperature range 5 to 70\,K and for the lowest hfs levels.  However, F.\,Lique kindly made available to us their entire results for upper energy  levels up to  $J_{\rm up} = 7$ prior to publication. The computation of the inverse rates, that is, the collisional excitation rate coefficients, assumed thermal balance. 

The region was devided into three parts:  the north, centred on N\,2, the middle, centred on SM\,1, and the southeast, centred on N\,6, where the N-nomenclature refers to the source designations by \citet{difrancesco2004}. However, only pixels with $\ge 3 \sigma$ integrated intensities in both the \ntwohp\,(3-2) and (6-5) lines were included. These regions, assumed to be spherical, are about 75\arcsec\ in size, hence neighbouring regions are overlapping and radii are 0.02\,pc ($6.75 \times 10^{16}$\,cm). We kept the turbulent velocity constant throughout the core, that is, $\upsilon_{\rm turb} = 0.25$\,\kms. After a number of initial trials spanning over a much larger parameter space, the ALI code was run on a finer grid around the most probable values, defined by $4.0 \le \log n$\,(\molh)$\,\le 7.0$ in steps of 0.2, $-11.0 \le \log X$(\ntwohp)\,$\le -8.0$ by 0.2 and $5\le T_{\rm kin}\le 50$\,K by increments of 1\,K, generating 11776 models. Among these, 2816 solutions were found that fit the observed intensity ratios within the observational errors. Using the individual line profiles, that is, both the shape and the absolute intensity, for both lines, \ntwohp\,(3-2) and (6-5) from the same position (pixel), the best-fit ALI solution was selected for that pixel. As best-fit-estimator, we used sums of the residuals, where the higher frequency spectra were weighted more strongly to compensate for their overall lower intensities. In other words, we minimized  $ \sum_{3\rightarrow 2} \left(T_{\rm obs}-T_{\rm ALI}\right)^2 + 10\times \sum_{6\rightarrow 5} \left(T_{\rm obs}-T_{\rm ALI}\right)^2$, resulting in a map of best-fit model profiles that, together with the observed ones,  are displayed in Fig.\,\ref{hfs_width}.

In Fig.\,\ref{aliresults}, the spatial distributions of the best-fit model parameters $N$(\molh), $T_{\rm kin}$, $n$(\molh) and $X$(\ntwohp) are shown. Values for the 10\asec\ pixels span two orders of magnitude in column density, $N$(\molh). For the star \ro, \citet{bohlin1978} determined the anomalous relation $N({\rm H\,I}+{\rm H_2})/E(B-V) = 15.4 \times 10^{21}$ atoms \cmtwo\,mag$^{-1}$. Since $R_{\rm V}=A_{\rm V}/E(B-V) = 5.5$,  the gas-dust relation for \ro\ reads $N({\rm H}_2)=1.4 \times 10^{21}$\,\av\,\cmtwo. Using this calibration also for the core, the ALI  results over 30\asec\ scales would imply the surprisingly low visual extinction, \av, through \roa\ of about 100\,mag or lower.  Average column densities of a few times \powten{22}\,\cmtwo\ 
are in accord with previous determinations. 

Kinetic gas temperatures are about 10-13\,K in the central regions, but 17-19\,K in the north, and16-19\,K in the southeast. Overall, these temperatures are lower than what has been determined before \citep[e.g.,][and references therein]{bergman2011b}.  
The relative abundance, $X$(\ntwohp)\,=\,$N$(\ntwohp)/$N$(\molh) appears to peak in the central part, where $\log X$(\ntwohp)\,$=-9.0\pm0.2$, whereas $X$(\ntwohp) is generally lower than \powten{-10} in the outer regions. There, derived overall gas volume densities, $n$(\molh), are around \powten{6}\,\cmthree, which would be in accord with what might be required by some models of the dust opacity (see below, Sect.\,4.4.2). Toward SM\,1, and on much smaller scales, \citet{friesen2014} estimated densities in excess of \powten{9}\,\cmthree.

As expected, optical depths in the centre of the (3-2) main line are high, explaining the low contrast between main and satellite components (Fig.\,\ref{aliresults}). In contrast, the (6-5) lines are optically thin, which is supported by their observed Gaussian line shapes. The figure also shows the distribution of the radial velocity residuals, $\Delta$v $= \upsilon_0 - \upsilon_{\rm LSR}$, where $\upsilon_0$ is the observed central velocity and \vlsr=+3.5\,\kms. Locally, velocity shifts are lower than 0.2\,\kms, which is much lower than the assumed turbulent velocity of the gas (0.6\,\kms). On much larger scales, $>40$\amin,  \citet{liseau1995} derived a NW-SE radial velocity gradient of about 1\,\kms\,pc$^{-1}$. For the limited extent of the region studied here, this has only
a negligible effect on the line transfer.

%
\begin{figure*}
  \resizebox{\hsize}{!}{
    \rotatebox{00}{\includegraphics{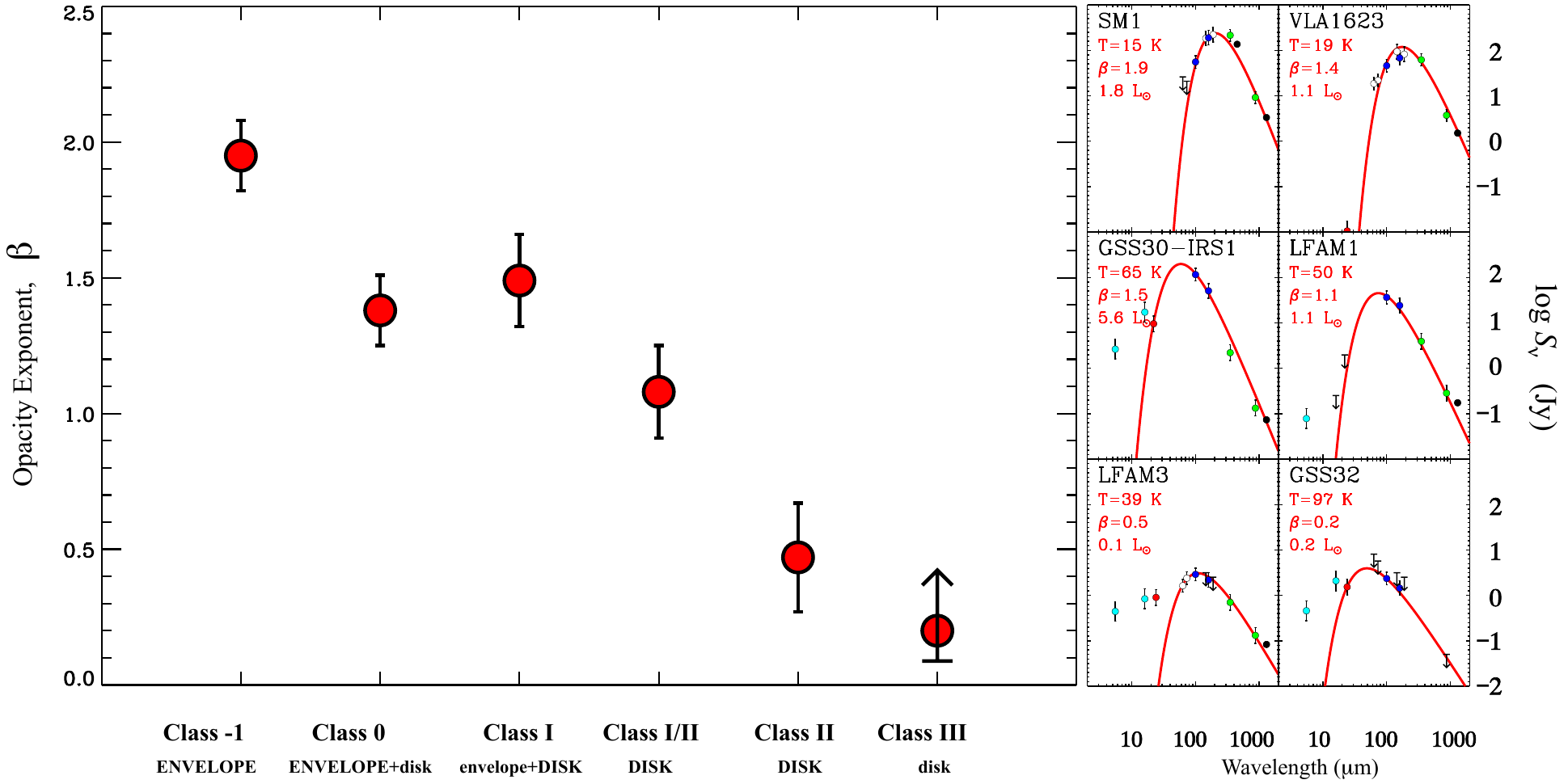}}
                        }
  \caption{{\bf Left:} Time evolution of the opacity exponent $\beta$, where $\kappa_{\nu} \propto \nu^{\,\,\beta}$, and the values are derived from the fits to the adjacent SEDs.
  {\bf Right:} SEDs of the point sources SM\,1, VLA\,1623, GSS\,30-IRS1, LFAM\,1, LFAM\,3 AND GSS\,32, belonging to different evolutionary classes (see Fig.\,\ref{N2H+3_2}). The $\beta$-values decrease from 2 for the starless dense dust core SM\,1 to 0.2 for the visible T\,Tauri star GSS\,32. Colour coding of the symbols is light blue: ISO-CVF 5.5\,\um\ and 16\,$\mu$m. Red: WISE 22\um\ or Spitzer 24\,$\mu$m. Blue:  {\it Herschel}-PACS 100\,\um\ and 160\,$\mu$m. White: PACS-spectra at 63, 73, 145 and 189\,$\mu$m. Green:  APEX 350 and 870\,$\mu$m. Black refers to literature data: \citet[][450\,\um]{wilson1999} and \citet[][1.3\,mm]{motte1998}.
         }
    \label{beta}
\end{figure*}


\subsection{Dust temperature, opacity, and mass}

\subsubsection{Point-source fitting of the SED}

The fluxes of several point sources in the field were extracted
and corrected for the underlying cloud emission.  Their SEDs were determined assuming modified blackbody emission, where the adjustable parameter $\beta$ is the power-law exponent of the frequency dependence of the flux in the Rayleigh-Jeans regime (see Appendix\,A). The fitting results are presented in Table\,\ref{beta_tab} and shown in Fig.\,\ref{beta}. These suggest a temporal evolution of the parameter $\beta$, in the sense that the FIR/submm SEDs become increasingly flatter with time. At early times, Class\,$-1$ and Class\,0, the FIR/submm emission is entirely dominated by the envelope, whereas at the later stages, the envelope emission has essentially vanished and is dominated by the disk or the remnants of it (Class\,III). A mixture of these two dust-emission components is exhibited by the objects in between (Class\,I and II).  The value of $\beta$ is presumably  dominated by the size distribution of the dust grains \citep[e.g.,][]{miyake1993,krugel1994}. 

To examine the $\beta$-dependence on the size $a$ of (spherical) grains,  we performed Mie-calculations \citep{min2003} using laboratory measurements for a few typical grain materials (Fig.\,\ref{beta_size}). This plot is qualitatively similar to Fig.\,4 of  \citet{testi2014} for young disks, but here, $\beta$ has been calculated for the wavelength region 200\,\um\ to 1.3\,mm instead, and it falls into the interval $0.0 \le \beta \le 2.5$. The power-law exponent  $p$  has been varied within the range $-5.0 \le p \le -2.0$, and where $dn(a) \propto a^{\,p}da$. As expected, the $\beta$-dependence on chemical composition decreases with increasing $p$, that is, grain size, and these larger grains are found around the Class\,II and III objects, viz. LFAM\,3 and GSS\,32, respectively, with $a$\,\gapprox\,100\,\um. In contrast, the very small grains are found predominantly around the probably youngest object, that
is, the starless clump SM\,1 (Class$-1$). Around the Class\,0 to I sources VLA\,1623, GSS\,30\,IRS1 and LFAM\,1, sizes would fall roughly within $0.1 \le a \le 10$\,\um.

%
\begin{table}[htdp]
  \caption{SED-fitting result for sources at different evolutionary stages.}
  \begin{tabular}{llcl}
    \hline
    \noalign{\smallskip}
    Source                     & $T_{\rm dust}$ [K]   & $\beta$                 &  Type         \\
    \noalign{\smallskip}
    \hline
    \noalign{\smallskip}
     SM1$^{a}$                  & $15 \pm 1$            & $1.95 \pm 0.13$       & Class\,-1             \\
     VLA1623                    & $19 \pm 2$            & $1.38 \pm 0.14$       & Class\,0              \\
     GSS\,30-IRS1       & $65 \pm 4$            &  $1.49 \pm 0.17$   & Class\,I              \\
     LFAM\,1                    & $51 \pm 6$            & $1.08 \pm 0.17$       & Class\,I              \\
     LFAM\,3            & $39 \pm 3$            & $0.47 \pm 0.21$         & Class\,II             \\
     GSS\,32            & $96 \pm 18$           & $\geq 0.2$                 & Class\,III    \\
    \noalign{\smallskip}
    \hline
  \end{tabular}
  \begin{list}{}{}
   \item[$^{a}$] {\small {\citet{ward-thompson1989} determined $T_{\rm dust}=15 \pm 5$\,K\\and $\beta=2.2 \pm 0.3$}}.
  \end{list}
  \label{beta_tab}
\end{table}

\begin{figure}
 \resizebox{\hsize}{!}{
    \rotatebox{00}{\includegraphics{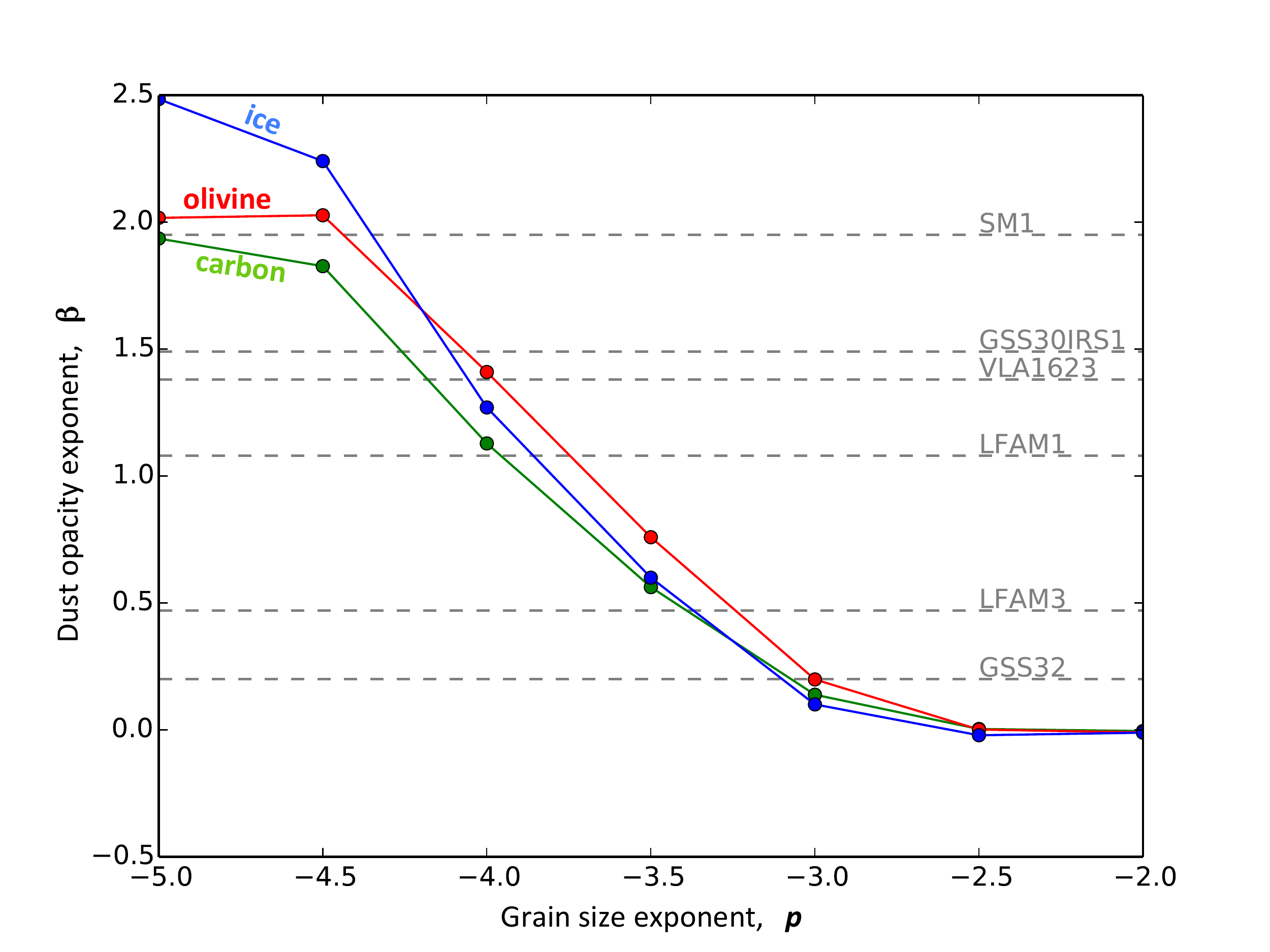}}
                        }
  \caption{Grain size exponent $p$, defined through $dn(a) \propto a^{\,p} da$ for spherical grains of size $a$, as a function of the grain opacity exponent $\beta$ for the wavelength range 200\,\um\ to 1.3\,mm, and where $\kappa_{\nu} \propto \nu^{\,\,\beta}$. The curves for three different materials are shown, viz. for amorphous ice \citep[\water, blue:][]{bertie1968}, amorphous silicates \citep[MgFeSiO$_4$, red:][]{jaeger1994} and amorphous carbon \citep[C, green:][]{blanco1991}.  The values of $\beta$ determined from observations are shown as dashed lines, referring to the sources of Table\,\ref{beta_tab} and identified to the right.
  }
  \label{beta_size}
\end{figure}


Therefore, the observed time dependence suggests that grain growth mainly occurs in the protoplanetary disks, with only little enrichment of large grains in the parental cloud. However, the precise dating of these events is difficult, but they probably fall within the range of some thousand years (starless cores, Class\,-1) to some megayears (near the ZAMS, Class\,III). If our observational results apply to currently adopted pre-main-sequence timescales {\citep[e.g.,][]{evans2009}, then $\beta(t)$ can be fit by a (broken) power law, $\beta \sim t^{\,-\,\gamma}$, where $\gamma(< 10^5\,{\rm yr}) \sim 0.04$ and $\gamma(\ge10^5\,{\rm yr}) \sim 0.5$ to 0.6, respectively. If $\beta$ scales inversely with grain size/mass, this may be consistent with stochastic processes in a turbulent medium \citep[see also][]{blum2004}. Apparently, very little happens up to some \powten{5} years. Thereafter, grain growth seems to occur at a faster rate.

Needless to say, it would be very valuable to extend this type of analysis to other star-forming regions, both for better statistics, but above all  also for the comparison among different environments. $\beta$-evolution around young sources has recently been found elsewhere, but generally only for a single object at a time or, in one case, for two sources \citep[][and references therein]{testi2014}. However, the very compactness ($0.05\,{\rm pc}\sim 10^4$\,AU) of an entire star-forming region  in combination with the multitude of objects of different type as discussed here, makes the present study of  \roa\ unique.

As for SM\,1, these results refer to source scales of some \powten{3}\,AU. On smaller scales, that is, a few \powten{2}\,AU, SM\,1 appears to have fragmented into smaller pieces \citep{nakamura2012}. This may be expected on the basis of current theoretical models of star formation \citep[][and references therein]{lomax2014}. However, using ALMA at higher resolution, \citet{friesen2014} found no evidence for substructures down to scales of some 10\,AU. 

\begin{figure}
 \resizebox{\hsize}{!}{
    \rotatebox{00}{\includegraphics{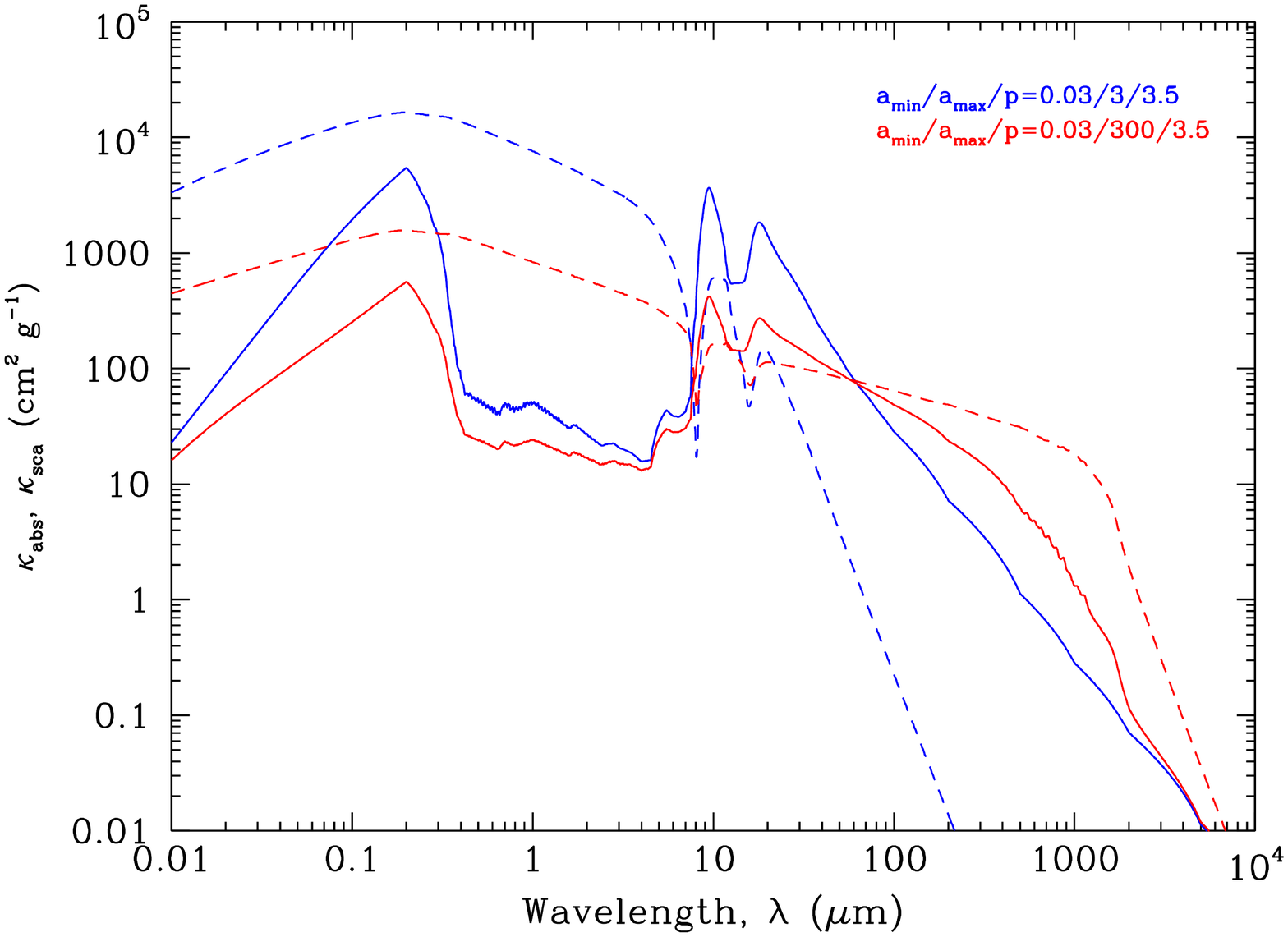}}
                        }
  \caption{Extinction curves used for Model\,I,  where the mass extinction coefficient is given by $\kappa_{\rm ext}(\lambda)=\kappa_{\rm abs}(\lambda)+\kappa_{\rm sca}(\lambda)$. Solid lines are for mass absorption coefficients $\kappa_{\rm abs}$ and dashes are for the scattering, $\kappa_{\rm sca}$. The curves are based on the optical constants provided by  \citet{jaeger1994} and by \citet{dorschner1995} for two grain populations. The adopted grain size distributions have an exponent $p=3.5,$ and minimal sizes are $a_{\rm min}=0.03$\,\um\ for both. The largest sizes are $a_{\rm max}=3$\,\um\ (blue) and  $a_{\rm max}=300$\,\um\ (red). }
  \label{bem_opac}
\end{figure}

\subsubsection{Extended dust emission}

The methods for estimating dust emissions from point and extended sources differ only slightly \citep[e.g.,][]{andre1993}, and in Appendix\,A, we present the analysis of our submm data assuming conventional methods, again for the purpose of comparison with other works. We did not, however, as is often done, assume a single dust temperature nor a single power law exponent $\beta$ of the frequency dependence of the dust opacity, nor a single normalization value at some fiducial frequency. Instead, we intend to exploit observations at, at least, three frequencies to determine the spatial distribution of the dust temperature, the dust optical depth, and the absolute grain mass opacity. However, from the observed spectral index $\alpha = \Delta \log I_\nu/ \Delta \log \nu$ for the wavebands 350\,\um--870\,\um\ and 870\,\um--1.3\,mm, respectively,  it is apparent that the APEX data are internally consistent, but appeared incompatible with the IRAM data (Fig.\,\ref{bocas}). Whereas $\alpha$ generally is $>2$ for the Saboca-Laboca maps, as one would expect for thermal emission in the Rayleigh-Jeans (RJ) regime, the index is $<2$ everywhere for the Laboca-IRAM data. We have to conclude that the relative calibration of the LABOCA and SABOCA maps is good, but that the LABOCA and IRAM data suffer from inter-calibration uncertainties, and the latter were therefore not used in our analysis (see Appendix\,A).

With $I_{\nu} $ obtained from observation, the right-hand side of Eq.\,\ref{surfden} contains two unknowns. For the interpretation of observations of molecular clouds, it has been common to assume ad hoc values for $\kappa_{\nu}$ and $T_{\rm d}$. In addition, one also assumes that the run of $\kappa$ with $\nu$ is given by a power law, with an exponent $\beta$, that is, $\kappa_{\nu} \propto\,\nu^{\,\,\beta}$. For thermal emission in the RJ-regime, $\beta$ would be related to the spectral index $\alpha$, given by $\beta = \alpha -2$, and for spherical Mie particles, $\beta$ takes values between 0 and 2 \citep[e.g.,][]{emerson1988}, but values lower than 0 and greater than 2 are possible when the proper physics of non-perfect crystalline material are included \citep{meny2007}. The use of the $\beta$-relation for mass estimates requires the normalization at some fiducial frequency \citep[cf.][]{hildebrand1983}. In many cases, the normalization value of the opacity incorporates an assumed $g2d$ of 100 \citep[][and references therein]{beckwith2000}.

The normalization is commonly taken from published works on dust opacity computations. However, by using these theoretical results directly, one circumvents the necessity of specifying $\beta$-values, which  can change from one frequency band to another. Instead, one would wish to determine how these might vary spatially across a cloud, which might tell us something about the cloud dust properties. Potential problems include noise fluctuations and calibration uncertainties, and gradients in density and/or temperature along the line of sight that may lead to non-convergence/inconsistent $\kappa$-values. Similar would apply to derived $\beta$-values, when one assumes that $\kappa(\nu)$ is given by a single power law of index $\beta$ \citep[e.g.,][]{juvela2013}. The accuracy of our results below is limited to twice the statistical error of the intensity $I_{\nu}$ at each frequency $\nu$. However, the largest uncertainty comes from the absolute value of the opacity.

As an example, we chose the work by \citet{ossenkopf1994}, which provides dust mass opacities for conditions thought to prevail in dense molecular clouds. Processes of coagulation are considered and the steady state is reached typically after some \powten{5} years, which is most likely younger than the ages of molecular  clouds. The initial state consists of bare grains, distributed in size as in the diffuse interstellar medium \citep[viz. ``MRN-distribution",][]{mathis1977}\footnote{Comparing mass absorption coefficients for MRN dust of \citet{draine2003} with those of \citet{ossenkopf1994} indicates some differences: \citet{draine2003} lists values of $\kappa_{350\,\mu {\rm m}}= 1.92$\,cm$^2$\,g$^{-1}$ and $\kappa_{850\,\mu {\rm m}}= 0.383$\,cm$^2$\,g$^{-1}$. The corresponding data by \citet{ossenkopf1994} are $\ge 3.64$\,cm$^2$\,g$^{-1}$ and $\ge 0.70$\,cm$^2$\,g$^{-1}$, respectively.}. With time, these grains grow in size by acquiring ice mantles of various thickness and opacities, between the wavelengths of 1\,\um\ and 1.3\,mm, are provided for a range in  environmental density, from $n$(H)\,=\,0 to \powten{8}\,\cmthree. 

Shown below, in Sect.\,4.2.4 (Fig.\,\ref{dustT}), are the results for fixing the opacity curve, that is, No.\,22 in Table\,\ref{ossen}. The distribution of the dust temperature exhibits an average value of 19\,K. Although temperatures are around 20\,K in most parts of the map, regions with $T_{\rm d} > 25$\,K exist along a north-south ridge, in addition to a couple of localized shocked regions, near VLA\,1623 and GSS\,30 (cf. Fig.\,\ref{classes}). The dust optical depth at 350\,\um, $\tau_{350\,\mu \rm m}$, is much lower than unity, justifying a posteriori the assumption of low optical depth. This is then of course also simultaneously justified for the LABOCA data ($\max \tau_{870\,\mu {\rm m}}=0.005$). 

A major drawback of the presented analysis is that it ignores the details of the processes that are heating the observed dust. In other words, the energy balance equation is not solved, which neglects the existence of gradients in density and radiation field strength, and hence temperature, in the source. These shortcomings are considered in the physical models described below, which are two 3D radiative transfer models that follow two different approaches. Regarding spatial coverage, these models are complementary, meaning that model I provides a larger, global view and model II focuses on the central regions of \roa.

\begin{table}
\caption{Dust masses for opacites from \citet{ossenkopf1994}.}             
\label{ossen}      
\begin{tabular}{clcccc}      
\hline\hline    
\noalign{\smallskip}             
$\kappa$&ice    &  $n(\rm H)$           & $\beta_{\rm theo}$            & Average                         & $100\,M_{\rm d}$      \\
curve &coating  & (\cmthree)            &  350-870\,\um         & $T_{\rm d}$ (K)         & (\msun)                       \\
\noalign{\smallskip}    
\hline                        
\noalign{\smallskip}                            
1       &no ice & 0                             & 1.84                          & \phantom{1}21.06                & 15.92                 \\
2       &               & \powten{ 5}           & 1.50                          & \phantom{1}45.88                & \phantom{1}3.28       \\
3       &               & \powten{ 6}           & 1.28                          & \phantom{1}81.44                & \phantom{1}0.81    \\
4       &               & \powten{ 7}           & 1.12                          & 107.06                          & \phantom{1}0.23       \\
5       &               & \powten{ 8}           & 1.05                          & 112.52                          & \phantom{1}0.14       \\
6       &thin           & 0                             & 1.89                          & \phantom{1}19.38                & 10.54                 \\
7       &               & \powten{ 5}           & 1.86                          & \phantom{1}20.16                & \phantom{1}7.42       \\
8       &               & \powten{ 6}           & 1.86                          & \phantom{1}20.10                & \phantom{1}5.79       \\
9       &               & \powten{ 7}           & 1.86                          &  \phantom{1}20.00               & \phantom{1}4.97       \\
19      &               & \powten{ 8}           & 1.86                          &  \phantom{1}20.24               & \phantom{1}4.66       \\
20      &thick  & 0                             & 1.89                          &  \phantom{1}20.36               & \phantom{1}7.69       \\
21      &               & \powten{ 5}           & 1.89                          &  \phantom{1}19.36               & \phantom{1}6.17       \\
22      &               & \powten{ 6}           & 1.89                          &  \phantom{1}19.43               & \phantom{1}5.52       \\
23      &               & \powten{ 7}           & 1.89                          &  \phantom{1}19.44               & \phantom{1}5.41       \\
24      &               & \powten{ 8}           & 1.92                          &  \phantom{1}18.91               & \phantom{1}5.61       \\
\hline                                   
\end{tabular}
\end{table}

\begin{figure*}[ht]
 \resizebox{\hsize}{!}{
    \rotatebox{00}{\includegraphics{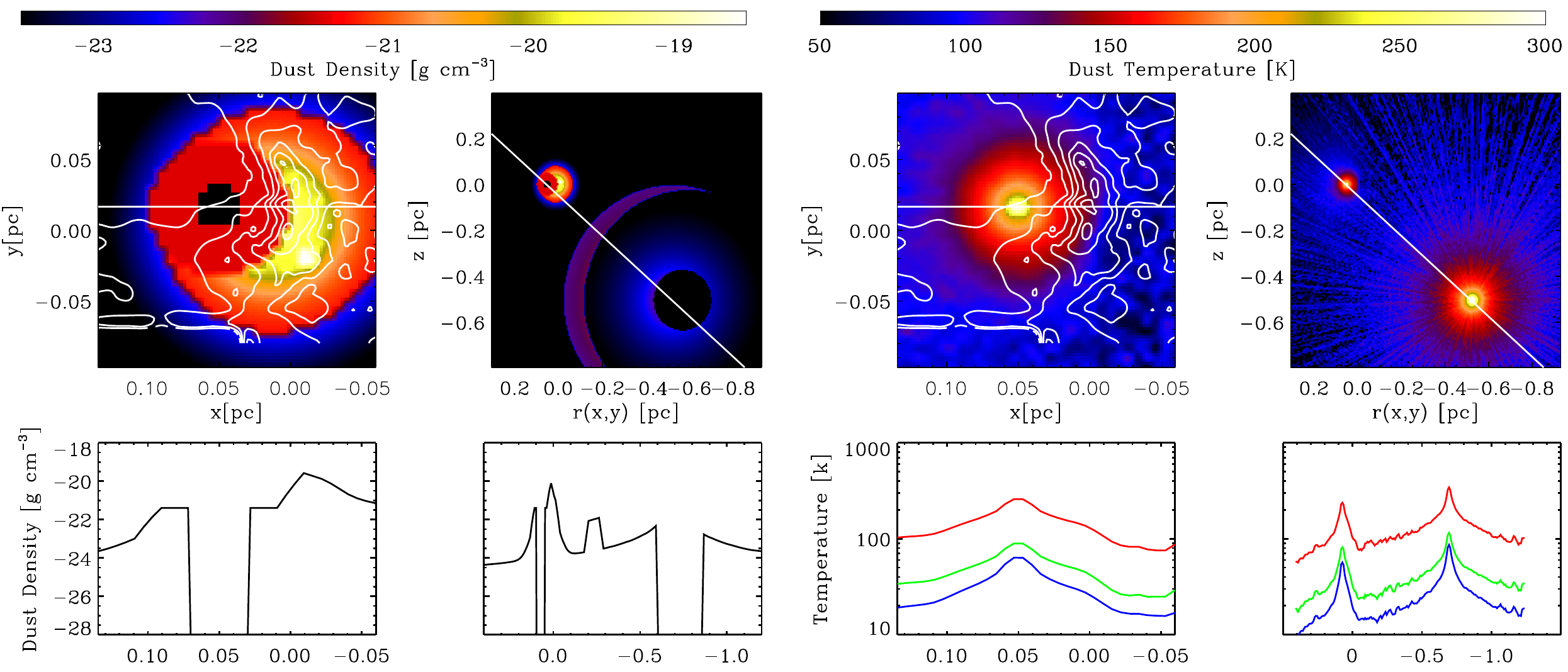}}
                        }
  \caption{{\bf Left panels:} Density distribution of model\,I. With the logarithmic colour bars on top, a 2D distribution is shown to the left that lies in a plane containing the star S\,1 and the centre of the main core SM\,1, and to the right the entire model, also containing HD\,147889. The panels below show the cuts along the straight white lines. White contours refer to the C$^{18}$O\,(3-2) measurements by \citet{liseau2010}.
{\bf Right panel:} Resulting temperature distribution of model\,I, with similar format as for the density distribution.  The different colours in the frames below refer to the temperatures of the various dust constituents - Red: Very small grains (VSGs) and PAHs. Green: grains with sizes between 0.03 - 3\,\um. Blue: grains whose sizes extend up to 300\,\um.
   }
  \label{BL_model_obs}
\end{figure*}

\begin{figure*}[ht]
 \resizebox{\hsize}{!}{
    \rotatebox{90}{\includegraphics{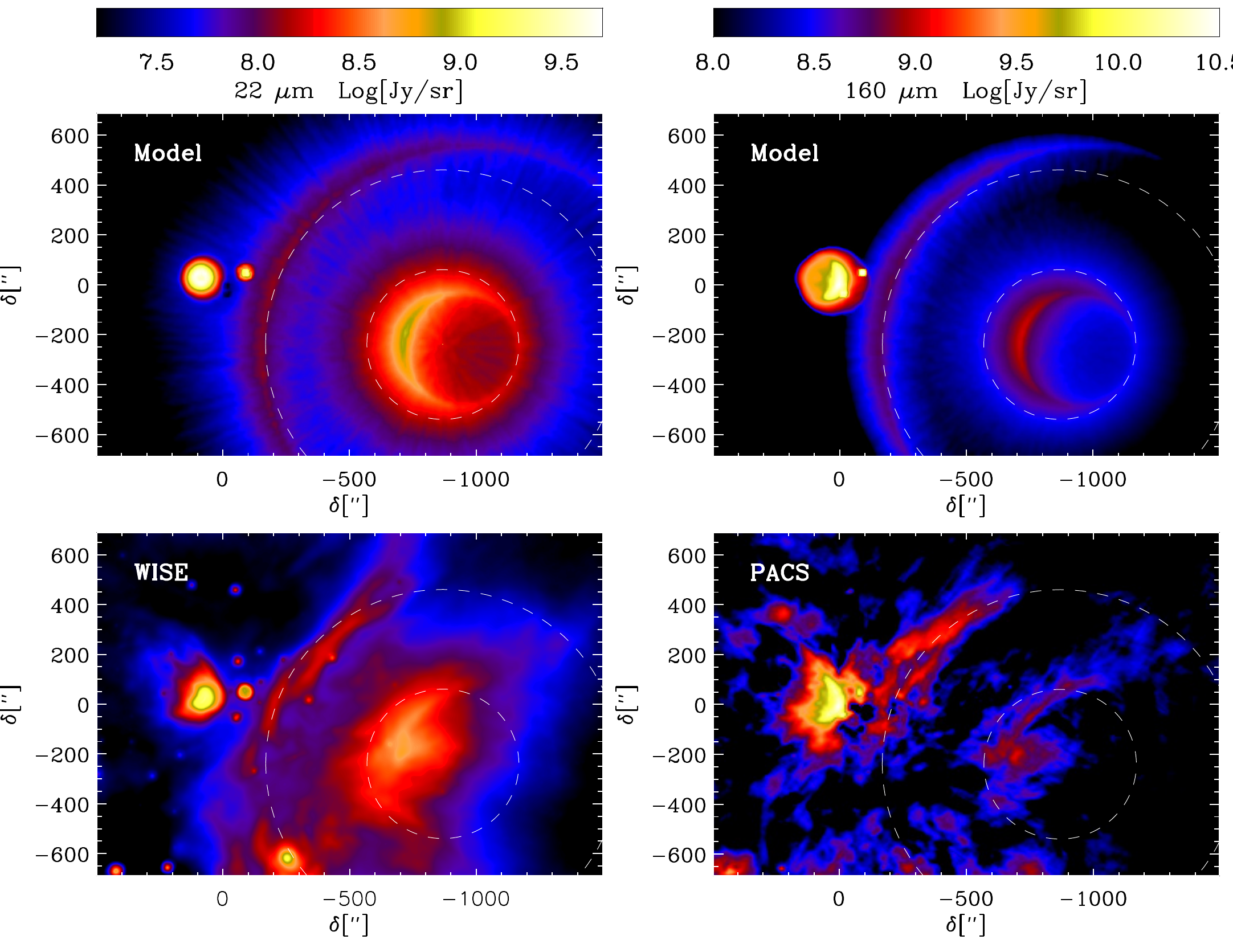}}
                        }
  \caption{Images of the full 3D dust model\,I at 22 \um\ and 160 \um\ for the entire \roa\ core.  {\bf Upper panel:}  Synthetic images at 22 \um\ and 160 \um. {\bf Lower panel:} The observations with WISE (22 \um, left) and PACS (160 \um, right).
     }
  \label{bem_total}
\end{figure*}

\begin{figure*}[htbp]
  \resizebox{\hsize}{!}{
     \rotatebox{90}{\includegraphics{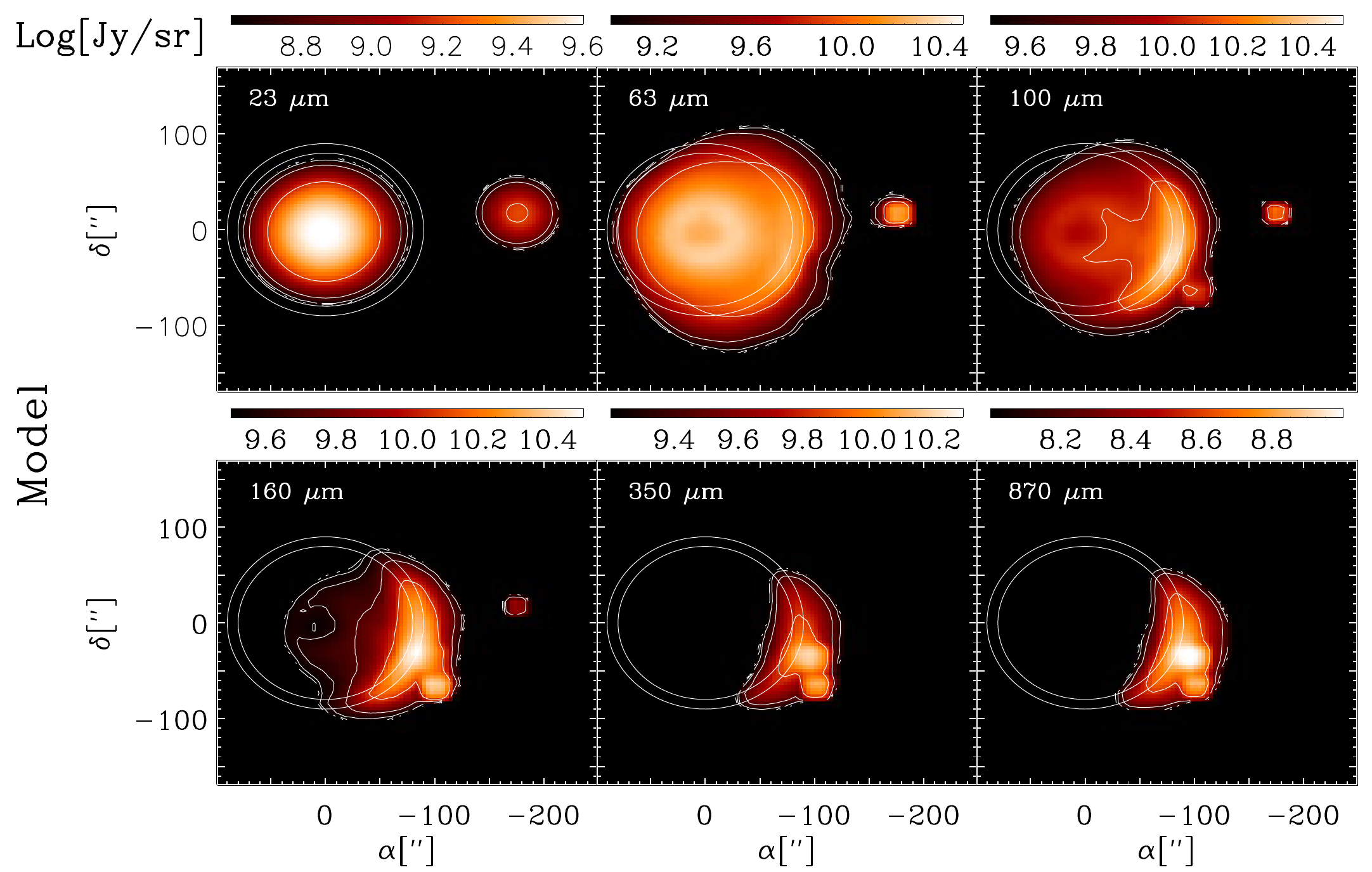}}
  }
  \resizebox{\hsize}{!}{
     \rotatebox{90}{\includegraphics{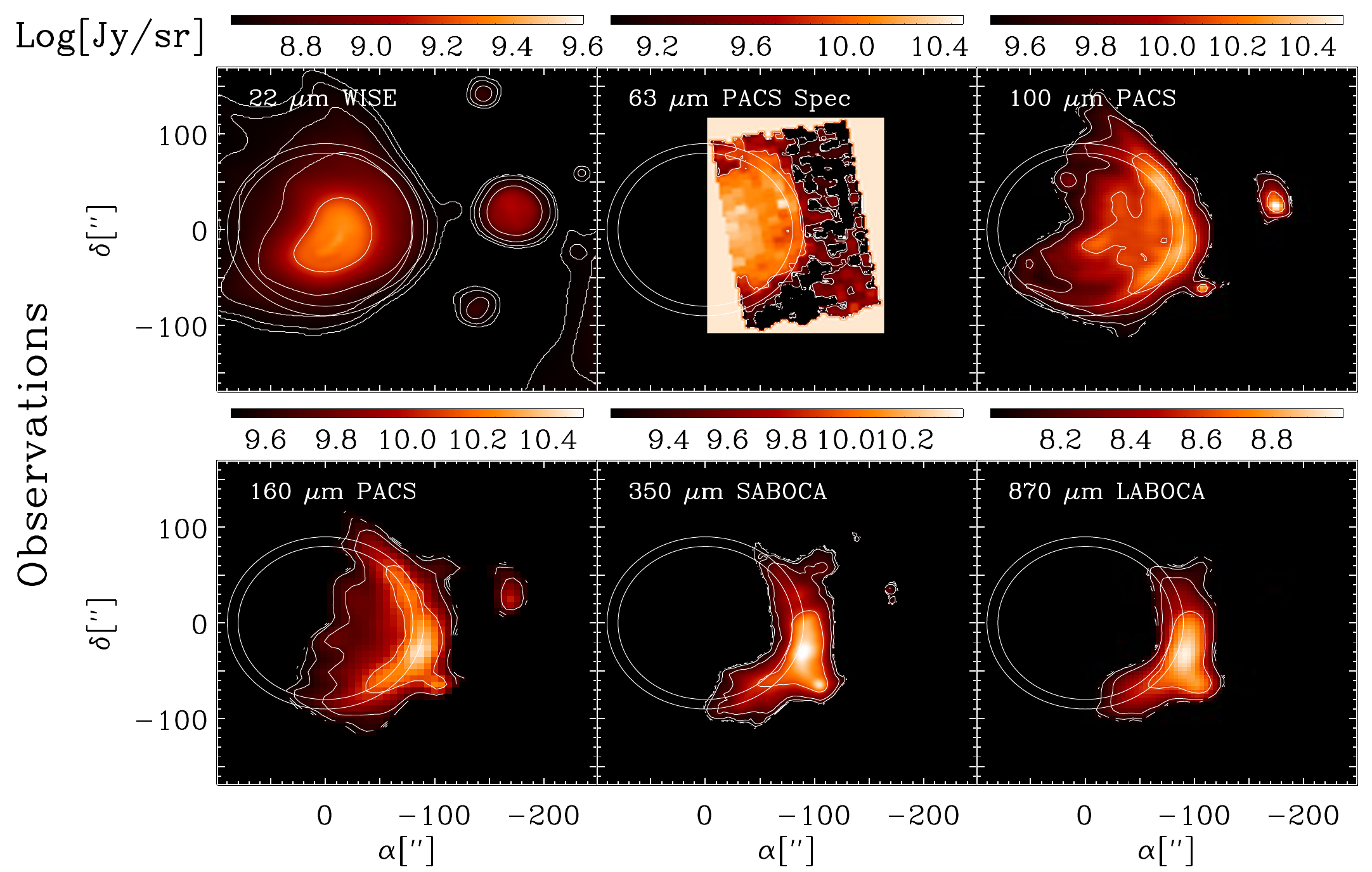}}
  }
  \caption{{\bf Top panels:} Synthetic  images of \roa\ that were generated from the results of model\,I.  Continuum wavelengths are 22, 63, 100, 160, 350, and 870\,\um. The logarithmic scales are in units of Jy\,sr$^{-1}$ and given along the colour bars atop each frame.  {\bf Bottom panels:} The displayed observations are taken from WISE (22 \um), {\it Herschel}-PACS (63, 100 and 160 \um), and APEX (350 and 870 \um). }
  \label{ima_obs_model}
\end{figure*}


\begin{figure*}
 \resizebox{\hsize}{!}{
    \rotatebox{00}{\includegraphics{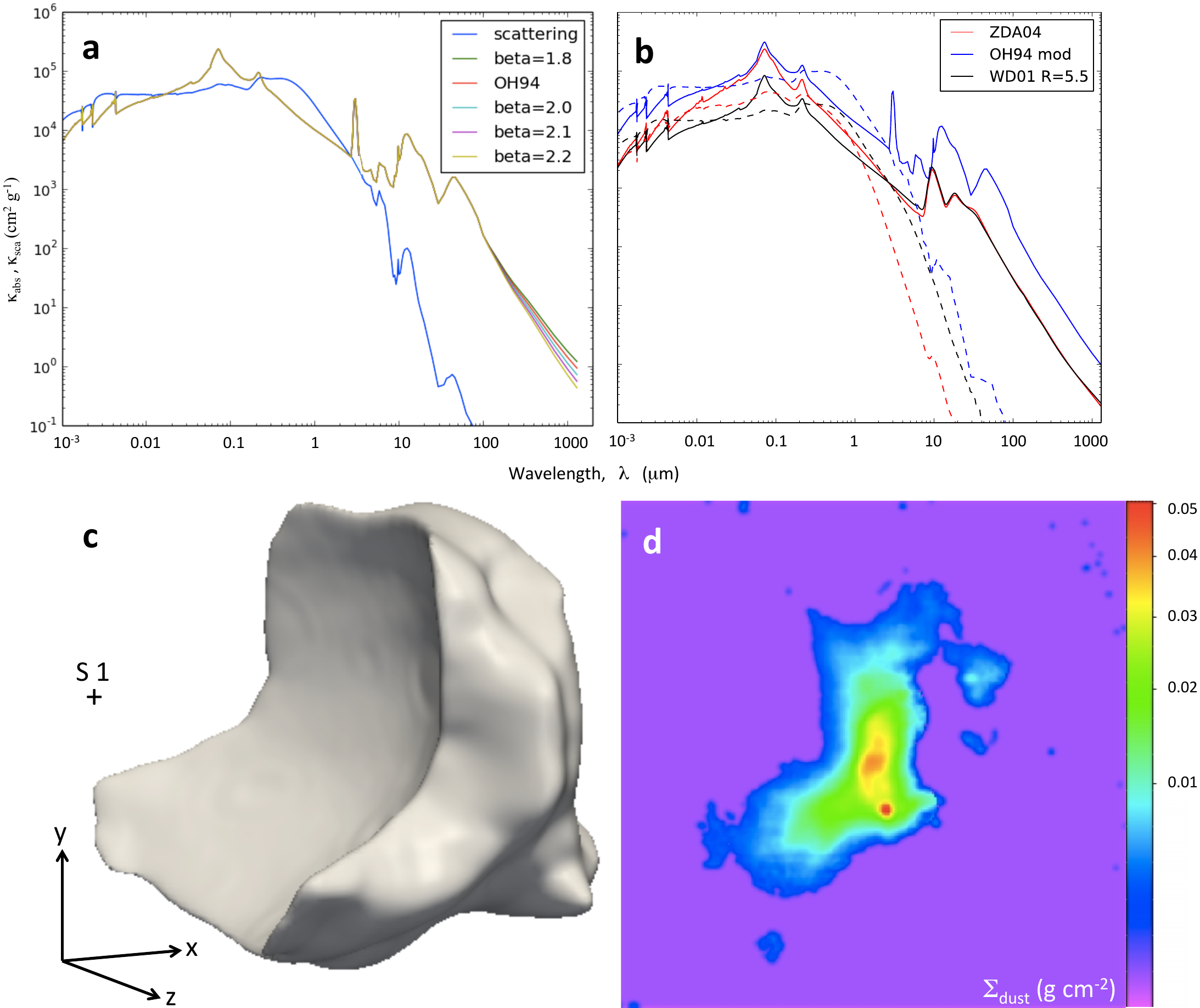}}    
                        }
  \caption{{\bf a.} Modification of the extinction curves of \citet{ossenkopf1994} (OH94) for various grain sizes and $\beta$-values (see the text). The extinction coefficients $\kappa_{\rm abs}$ and $\kappa_{\rm sca}$ of model Milky Way\,A for $R_{\rm V}=5.5$ of \citet{weingartner2001} match at 1\,\um. 
  {\bf b.} The extinction curves of \citet[][ZDA04 in red]{zubko2004} and \citet[][WD01 in black]{weingartner2001}. The modified curve of  \citet{ossenkopf1994} is shown in blue (OH94 mod). Solid lines depict mass absorption, dashed lines scattering. 
  {\bf c.} Structure of model\,II, where the isodensity surface corresponds to 20 percent of the highest density. The surface encloses approximately 30 percent of the total mass of the core. The line of sight (LOS) is along the negative $z$-axis, i.e. the axis pointing towards the observer. In the sky, right ascension is along the negative $x$-axis and declination along the positive $y$-axis. The dominating heating source S\,1 is situated near the focus of the paraboloid low-density region. 
  {\bf d.} The surface density distribution of the dust, $\Sigma_{\rm dust}$ in  g\,\cmtwo, for model\,II.3, i.e. for the extinction curve \citep[][$R_{\rm V}=5.5$ with PAHs]{weingartner2001}.
   }
  \label{TL_model}
\end{figure*}

\begin{figure*}
\centering
 \includegraphics[width=11.25cm]{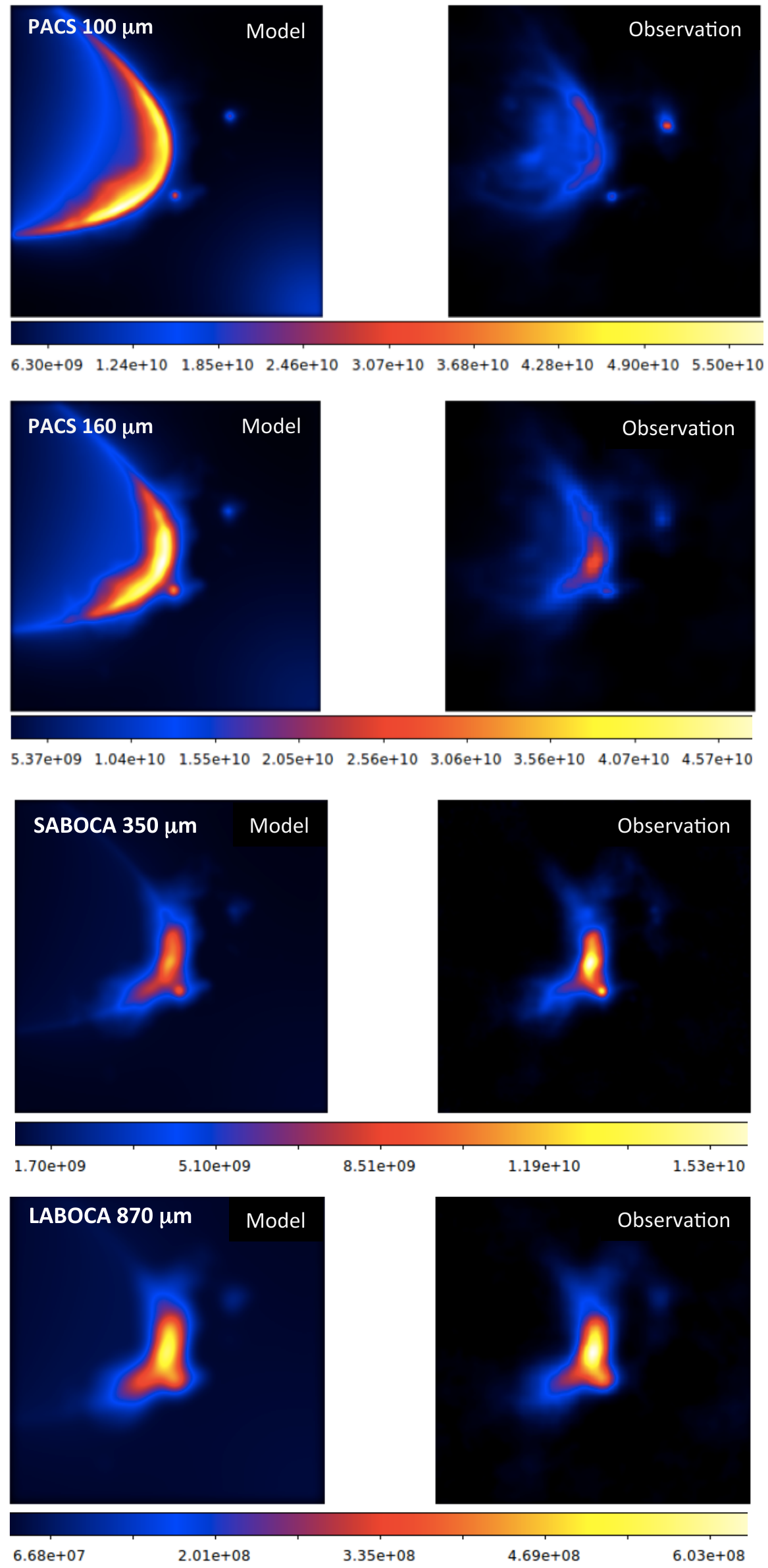}
    \caption{Results from model II.1, comparing modeling results (left) with the observations (right). From top to bottom PACS\,100\,\um, PACS\,160\,\um, SABOCA\,350\,\um,\ and LABOCA\,870\,\um, where intensity units are Jy\,sr$^{-1}$.  Including the PAH opacities did not improve the models at the FIR wavelengths, where the models over-predict the intensities observed by {\it Herschel}. In contrast, the agreement between models and APEX observations in the submm is very satisfactory, within better than 20\%.}
  \label{TL_model_obs}
\end{figure*}


\subsubsection{3D  dust modelling: Model I}

Model\,I  uses the code RADMC-3D\footnote{\small{\texttt{http://www.ita.uni-heidelberg.de/dullemond/software/\\radmc-3d/}}} developed by C.P. Dullemond and benchmarked by \citet{pinte2009}. The model consists of a low-density widespread cloud medium, the high-density core \roa\ with the dust clump SM\,1, the two point sources  VLA\,1623 and GSS\,30, and the two early-type stars S\,1 (B\,4) and HD\,147889 (B\,2.5), each with its own PDR bubble. These are the main energy sources in the model, providing the heating of the dust. The stellar SEDs are based on Kurucz ATLAS\,9 model atmospheres with the parameters of Table\,\ref{mod_tab}.

\begin{table}[htdp]
  \caption{Stellar parameters used in model\,I}
  \begin{tabular}{lcc}
    \hline
    \noalign{\smallskip}
    Parameter                   & S\,1   & HD\,147889   \\
    \noalign{\smallskip}
    \hline
    \noalign{\smallskip}
    $M$ (\msun)                         &   6               & 12 \\
    $R$ (\rsun)                 &   3              &   7 \\
    $T_{\rm eff}$       (K)             & 17\,000          & 22\,000 \\
    $\log g$ (in dyn \cmtwo)&  4.0           &  4.0 \\
    $\upsilon_{\rm turb}$ (\kms)&  2.0    &  2.0 \\
    \noalign{\smallskip}
    \hline
  \end{tabular}
  \label{mod_tab}
\end{table}

In detail,  the model assumes a high-density core of radius 0.045\,pc that is embedded in a low-density cloud. The early-type stars S\,1 and HD\,147889 generate two PDRs, viz. west and east of the dense core, with respective radii of 0.06\,pc and 0.5\,pc. The bright B2 star HD\,147889 is situated more than half a parsec away to the southwest and behind the cloud \citep{liseau1999}. Together, these stellar radiation sources are heating the dust of \roa\ from the outside. By comparison, internal heating sources are essentially negligible. 

The exploited grain opacities are shown in Fig.\,\ref{bem_opac}, and in \roa, individual sources contribute to the total modelled dust mass of $10.4\times 10^{-2}$\,\msun\ as follows: the core with 7, SM\,1 with 1.7, VLA\,1623 with 0.9, and GSS\,30 with 0.8 in units of \powten{-2}\,\msun.  A match to the PDR around HD\,147889 assumes 0.04\,\msun\ of dust (see Fig.\,\ref{BL_model_obs}). Based on C$^{18}$O and other molecules (Paper\,II, in preparation), the density profile is given by $\rho\,\propto\,r^{-1.5}$ for \roa\ globally and  by $\rho\,\propto\,r^{-2}$ for SM\,1 locally.

In the model, the stellar object S\,1 \citep[B5]{chini1981}\footnote{This is a binary \citep[B4+K,][]{gagne2004}. The magnetic B star \citep{andre1991} is also a bright source of X-ray emission \citep{hamaguchi2003, gagne2004}.}  is situated off-centre  in a low-density bubble of radius 0.06\,pc that is filled with $7 \times 10^{-4}$\,\msun\ of 0.1\,\um\ sized carbon particles\footnote{For $M/L=4.6 \times 10^{-3}$ of an equivalent single star, the blow-out grain size is larger than 50\,\um, and radiation pressure would empty the 12\,000\,AU region of carbonaceous particles of size 0.1\,\um\  within 3\,500 years. The likely existence of an intense stellar wind, exerting additional outward momentum, would shorten this timescale even further.} \citep{dorschner1995}. These small particles scatter the short-wavelength radiation extremely well, and this is needed to account for the observed intensity distribution in the near- and mid-IR. However, some finite, non-zero filling of the bubble is also required for numerical reasons. For an assumed $g2d=100$, the total mass from model\,I of the entire \roa\ core is 23.7\,\msun\ (Fig.\,\ref{bem_total}) .

As already mentioned, we used RADMC-3D to solve for the temperature distribution, which is shown along two different cuts on the right-hand side of Fig.\,\ref{BL_model_obs}. Obviously, the grain temperature also depends on its size, and the distributions for the three groups of grains are shown individually. For  \um- and mm-size grains, temperatures of the dust amount to 10-30\,K near the dense core SM\,1. These model results can be used to generate synthetic images of \roa. These are shown in Fig.\,\ref{ima_obs_model}, together with the observations at six wavelengths, from the mid-infrared to the submm. Evidently, model\,I reproduces the observed intensity distributions very well,  in particular the large-scale features.

\begin{table*}
\caption{Assumed submm/mm dust opacities, normalized to $g2d=100$.}             
\label{opacs}      
\begin{tabular}{lcccccl}      
\hline\hline    
Opacity reference               &        $\kappa_{350\,\mu {\rm m}}$    &  $\kappa_{870\,\mu {\rm m}}$ &  $\kappa_{1.3\,{\rm mm}}$   &       $M^{\,a}$               &       $(\kappa\,M)/(\kappa_0\,M_0)$&  Note  \\
                                        &       (cm$^2$\,g$^{-1}$)                      & (cm$^2$\,g$^{-1}$)               &  (cm$^2$\,g$^{-1}$)          &       (\msun)         &       350,\,\,\,870,\,\,\,1300\,\um           &       \\
\noalign{\smallskip}    
\hline                        
\noalign{\smallskip}
{\bf 1D-models}                 &                                                       &                                                &                                               &                               &                                                         &       \\
\noalign{\smallskip}    
This work       ($M_0,\,\kappa_0$)&     11.78                           & 2.148                                    & 0.9621                                       &\phantom{1}5.5 &   1.0,\,\,\,1.0,\,\,\,1.0                               &        $T$=10 - 50\,K, OH No.\,22, $\beta=1.89$  \\
\citet{johnstone2000}   &       $\cdots$                                        & 1.0     $^{a,\,b}$                               & $\cdots$                             &\phantom{1}8.1 (5.2) & \hspace{0.3cm}0.7\,(0.4)                &       $T$=const=20\,K ($T$=13 - 42\,K)  \\
\citet{andre2007}               &       $\cdots$                                        & $\cdots$                                 & 0.5$^a$                              & \phantom{1}7.5   &   \hspace{1.5cm}0.5                  &       $T$=const=12 or =20\,K \\
\noalign{\smallskip}    
{\bf 3D-models}                 &                                                       &                                                &                                               &                               &                                                       &       \\              
\noalign{\smallskip}    
Model\,I $^{c}$                 &\phantom{1} 2.87                               & 0.406                                   & 0.1846                                        &       10.4                    &   0.5,\,\,\,0.4,\,\,\,0.4                               &  MgFeSi grains for $a_{\rm max}=3$\,\um \\
Model\,II.1      $^{d}$         &       11.17                                   & 1.825                                    & 0.8206                                       &\phantom{1}7.3 &   1.3,\,\,\,1.1,\,\,\,1.1                               &  Modified OH No.\,22, i.e., $\beta=1.99$ \\       
Model\,II.2 $^{e}$              &\phantom{1} 2.05                               & 0.376                                    & 0.1886                                       &        36.7                    &  1.2,\,\,\,1.2,\,\,\,1.3                              &  Mass for $g2d=100$      \\
Model\,II.3     $^{f}$          &\phantom{1} 2.00                               & 0.393                                    & 0.2073                                       &        36.2                    &  1.1,\,\,\,1.2,\,\,\,1.4                              &  $R_{\rm V}=5.5$ and PAHs \\
\noalign{\smallskip}           
\hline                                 
\end{tabular}
    \begin{list}{}{}
    \item[$^{a}$]  For the \roac\  with assumed gas-to-dust mass ratio of $g2d=100$.
    \item[$^{b}$]  Refers to 850\,\um.
    \item[$^{c}$]  Opacities from \cite{jaeger1994,dorschner1995}.
    \item[$^{d}$]  Modified Curve No.\,22 of \citet{ossenkopf1994} using submm-$\beta=1.99$, but normalized at 100\,\um, i.e $\beta_{350-100\,\mu{\rm m}}=2.18$.
    \item[$^{e}$]  $M$\,=\,59.1\,\msun:  unmodified BARE-GR-S assumes $g2d=161$ \citep{zubko2004}.
    \item[$^{f}$]   $M$\,=\,38.0\,\msun: $R_{\rm V}=5.5$ curve of \citet{weingartner2001} that includes PAHs assumes $g2d=105.1$.
    \end{list}
\end{table*}

\subsubsection{3D  dust modelling: Model II.}

\noindent
{\it Thermal equilibrium models}
\\
\\
The modelling follows the procedure outlined by \citet{juvela2011}, which starts by deducing the 350\,\um\ optical depth distribution for a fixed $\beta=2$ from the observed APEX (350\,\um\ and 870\,\um) maps. The optical depth map is used to set the column density distribution of the model. However, the line of sight (LOS) density distribution is not directly constrained by the observations. We assumed a Gaussian distribution with a FWHM of 0.05 pc towards the densest part of the core, increasing up to 0.07 pc towards the low-density regions. With these values the LOS extent of the core is similar to its plane-of-sky dimensions. The resulting 3D density distribution was further modified to include a cavity around the S\,1 star. The cavity was modelled as a paraboloid with the star S\,1 at the focus. Inside the cavity the density was set to $10^{-5}$ times the highest density of the model core. The models represent a box with a size of 0.225 pc at a distance of 120 pc and use $256^3$ cells, yielding a resolution of approximately 180 AU, or \asecdot{1}{5}. The three-dimensional density distribution and the plane-of-sky surface density distribution are illustrated in the bottom left and right panels of Fig.\,\ref{TL_model}, respectively.

The model was embedded in an interstellar radiation field with intensity three times that of the solar neighbourhood. However, irradiation of the core was dominated by the individual stars, mainly S\,1 and HD\,147889. The stellar object S\,1 was modelled as a 16\,000\,K, $L=1\,100$\,\lsun\ blackbody, while HD\,147889 was given a temperature of 20\,000\,K and luminosity $L=4\,500$\,\lsun. Because HD\,147889 is more than 0.5\,pc away from the dense parts of the core compared to less than 0.1\,pc for S\,1, the latter is the main heating source in most parts of the model core. In addition, the model included two deeply embedded stellar objects with luminosities of approximately 2\,\lsun, but these do not have any significant effect on the radiation field except in their immediate surroundings. The radiative transfer calculations were made using the CRT program \citep{lunttila2012}. The program calculates the dust temperature taking into account the dust self-absorption and heating, and integrates along the line of sight to produce maps of dust FIR emission. For comparison with observations, the synthetic maps were convolved with Gaussian kernels corresponding to the APEX beam sizes.

We have calculated grids of models with varying dust properties and density scaling, but keeping the same radiation sources and the three-dimensional density distribution. In model II.1 we adopted the mass extinction from curve No.\,22 in Table\,\ref{ossen}, taken from \citet{ossenkopf1994}; however, the opacities are given only for wavelengths $\lambda \ge1.0$\,\um\ and no information is given about scattering properties. Therefore, for the shorter wavelength extinction, the model {\it Milky Way\,A} for $R_{\rm V} = 5.5 $ of \citet{weingartner2001} was used, matched to the former at the wavelength of 1.0\,\um. The scattering properties were obtained by assuming the same albedo and scattering asymmetry parameter as in the \citet{weingartner2001} $R_{\rm V} = 5.5$ model. To study the effect of $\beta$, in addition to the unmodified $\beta\approx 1.9$ dust, we used models where the extinction curve at wavelengths $\lambda \ge100$\,\um\ was modified by multiplying the \citet{ossenkopf1994} absorption cross-sections by $(\lambda/100\,\mu \rm m)^a$ with $a=-0.1$, 0.1, 0.2, and 0.3, corresponding approximately to $\beta=1.8$, 2.0, 2.1, and 2.2. The extinction curves are shown in Fig.\,\ref{TL_model}a and b. The peak dust surface density was varied in eight steps between 0.00501\,g\,\cmtwo\ and 0.0130\,g\,\cmtwo.

In model\,II.1, the best fit was obtained with $\beta=2.0$ and a peak dust surface density $\Sigma_{\mathrm{dust}}=0.010$\,g\,\cmtwo. Dust models with lower $\beta$ predict a too high 870\,\um\ to 350\,\um\ brightness ratio. Assuming a gas-to-dust mass ratio of 100, the mass of the main part of the core (defined as having at least 5\% of the peak surface density and corresponding roughly to the extent that was used for the 1D analysis in Sect. 4.2.2) is 7.3\,\msun. The mass included in the whole simulated region is approximately 20 percent higher. Although the best models fit the observed 350\,\um\ and 870\,\um\ intensity within approximately 20 percent, none of the models is successful in reproducing the 100\,\um\ and 160\,\um\ \emph{Herschel}-PACS observations. The calculated surface brightness is at least a factor of two higher than observed. In particular, it is difficult to simultaneously produce the relatively high brightness at 350\,\um\ and the low observed intensity in the PACS bands. Figure\,\ref{TL_model_obs} shows the comparison between the simulated maps from the best-fit model\,II.1 and the \emph{Herschel}-PACS 100\,\um\ and 160\,\um, and the SABOCA 350\,\um, and LABOCA 870\,\um\ maps. To study the effects of numerical resolution on the results, we increased the resolution by a factor of four, that is, to a cell size of 0.0002\,pc = 45\,AU. The differences between lower and higher resolution were lower than 10\% everywhere and lower than 2\% where the processes were active. In the following, we therefore ran the models with the lower resolution to keep the computations tractable.

In model\,II.2 we have chosen BARE-GR-S of \citet[][Table\,7]{zubko2004} as our basic dust opacities\footnote{These opacities do not require any ad hoc adjustments of the silicon or other abundances \citep[see the discussion by][]{draine2009}.}. As with model II.1, we ran simulations with modified dust models with different values of $\beta$. The modifications of the extinction law were similar to model II.1, except that the changes were applied at wavelengths $\lambda>20$\,\um. In addition to the standard $\beta\approx 2.0$ model we used $\beta=1.5$, 1.8, 2.1, and 2.2.  The density scaling was varied with nine steps corresponding to peak dust surface densities (at \asecdot{1}{5} scale) between 0.0197 g\,\cmtwo and 0.0985 g\,\cmtwo. As the \citet{zubko2004} dust models have much lower sub-mm opacities than the dust used in model\,II.1, in both cases the models span a similar range of 350\,\um\ optical depths. The best-fit II.2 model has a peak dust surface density $\Sigma_{\mathrm{dust}}=0.0498$\,g\,\cmtwo\ and $\beta=2.1$. Modified dust with a steeper sub-mm extinction law matches the 870\,\um\ to 350\,\um\ brightness ratio better than the original dust model in this case as well. Using the same \av-to-gas mass ratio as in the BARE-GR-S model of \citet{zubko2004} for our modified $\beta=2.1$ dust, the total mass of the best-fit model is 59.1\,\msun, where a $g2d=161$ has been assumed. 

\

In all models, the amount of dust in the cavity is very low, resulting in a visual extinction between the exciting source S\,1 and the inner edge of the cavity of $A_{\rm V}=3 \times 10^{-3}$\,mag. The extinction toward the observer is less well constrained, but model\,II.2 gives  $A_{\rm V}$ \lapprox\,15\,mag, with similar values for the others.  This value compares very well with the \av\ = 12.5\,mag determined by \citet{ward-thompson1989} from the \water-ice absorption feature at 3.1\,\um, and the range \av\ = 11 to 13\,mag, for $R_{\rm V}=5.5$, based on the X-ray analysis  by \cite{vuong2003}. 
\\
\\
\noindent
{\it Non-equilibrium models}
\\
\\
Models II.1 and II.2 assume that the dust grains are in thermal equilibrium with the local radiation field. However, absorption of a single high-energy photon can heat a very small dust grain to a high temperature. Absorbed energy is therefore reradiated more in the infrared than in the sub-mm \citep[see, e.g.,][]{draine2003}. To examine whether an accurate calculation of the emission from very small grains changes the results, model\,II.3 uses the dust model {\it Milky Way\,A} for $R_{\rm V} = 5.5 $ of \citet{weingartner2001} but does not assume thermal equilibrium. Because these calculations are much more time-consuming, only three different density scalings were used with this dust model. At wavelengths $\lambda>200$\,\um\ the calculated emission from the best-fit model is almost identical to the best-fit model\,II.2. At 100\,\um\ and 160\,\um, model\,II.3 produces slightly lower fluxes as shown in Fig.\,\ref{TL_SEDs}, but the discrepancy between the model and \emph{Herschel}-PACS observations remains large. The best-fit model has a peak dust surface density of  $\Sigma_{\mathrm{dust}}=$ 0.0491 g\,\cmtwo\ and, using a gas-to-dust mass ratio of 105.1, a mass of 38.0\,\msun.

The line-of-sight mass-averaged dust temperature, $T_{\rm dust}$, and the 350\,\um\ optical depth, $\tau_{350\,\mu {\rm m}}$, from the best-fit 1D model (see Sect. 4.2.2) and the three 3D models II.1, II.2, and II.3 are shown in Fig.\,\ref{dustT}. The results from all three 3D models are similar. Their LOS optical depths agree within 20 percent and the calculated dust temperatures only differ by approximately 1 to 2 K. The large differences in calculated masses are caused mostly by the widely varying sub-mm opacities and, to a smaller degree, by the gas-to-dust mass ratios of the dust models. The LOS average temperature through the dense core is between 18 and 20 K, but along the borders of the low-density cavity around the S\,1 star the temperature rises to 25--30 K. Results from the 1D model show much larger variations in temperature, although the average value is similar. The 350\,\um\ optical depths from the 1D analysis are approximately 30 \% lower.


\begin{figure}
 \resizebox{\hsize}{!}{
    \rotatebox{00}{\includegraphics{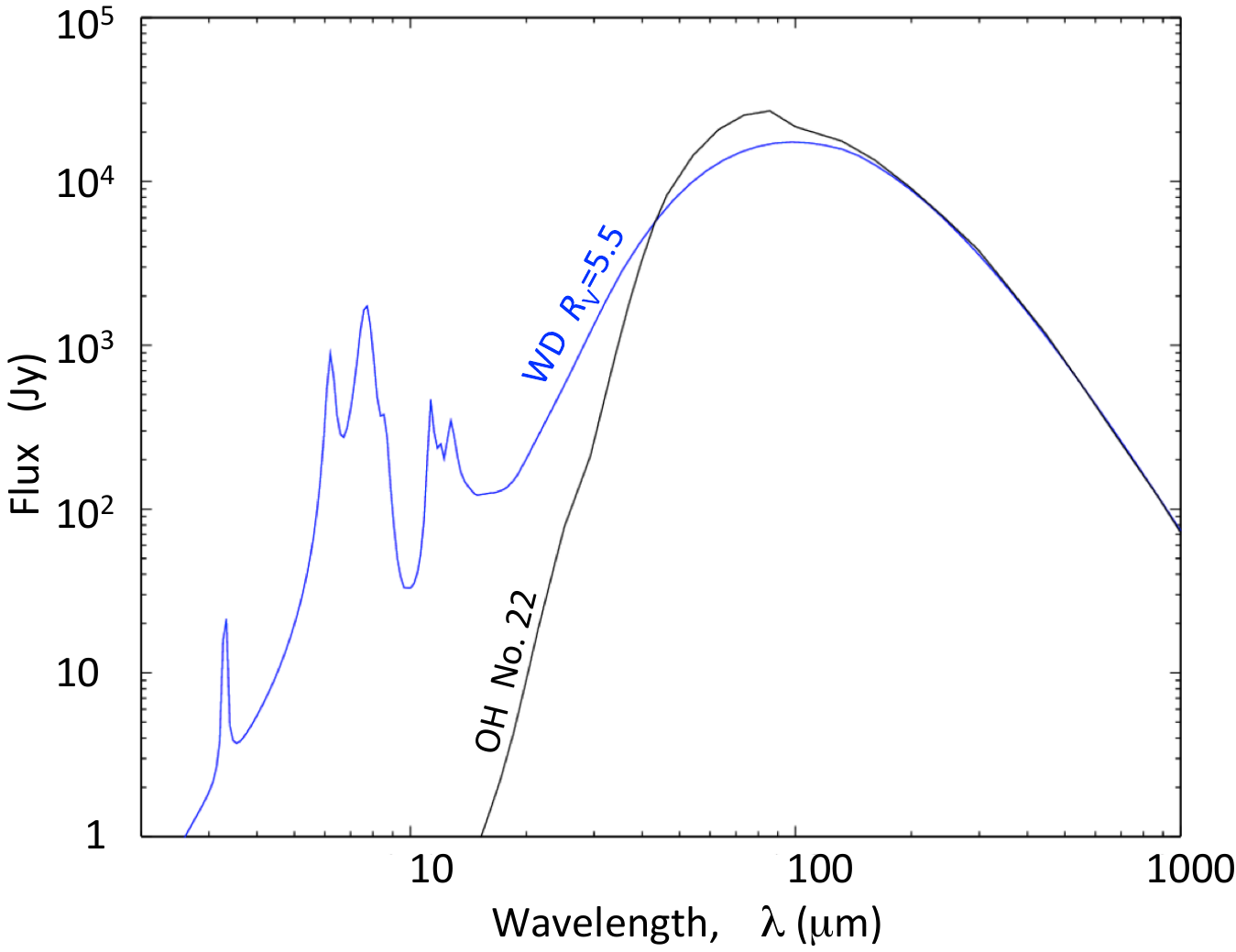}}
                        }
  \caption{SEDs of  \roa. For models II.1 (black) and II.3 (blue), the emerging specific intensity has been integrated over the core \roa\ that is displayed in Fig.\,\ref{TL_model}\,d. The thermal equilibrium model\,II.1 uses the extinction curve as
modified by \citet{ossenkopf1994}. The non-equilibrium model\,II.3 uses that of  \citet{weingartner2001} for $R_{\rm V}=5.5$, which includes the PAHs, and where by comparison, the 100 and 160\,\um\ fluxes are reduced. 
   }
  \label{TL_SEDs}
\end{figure}

\begin{figure*}
\centering
   \includegraphics[width=15.6cm]{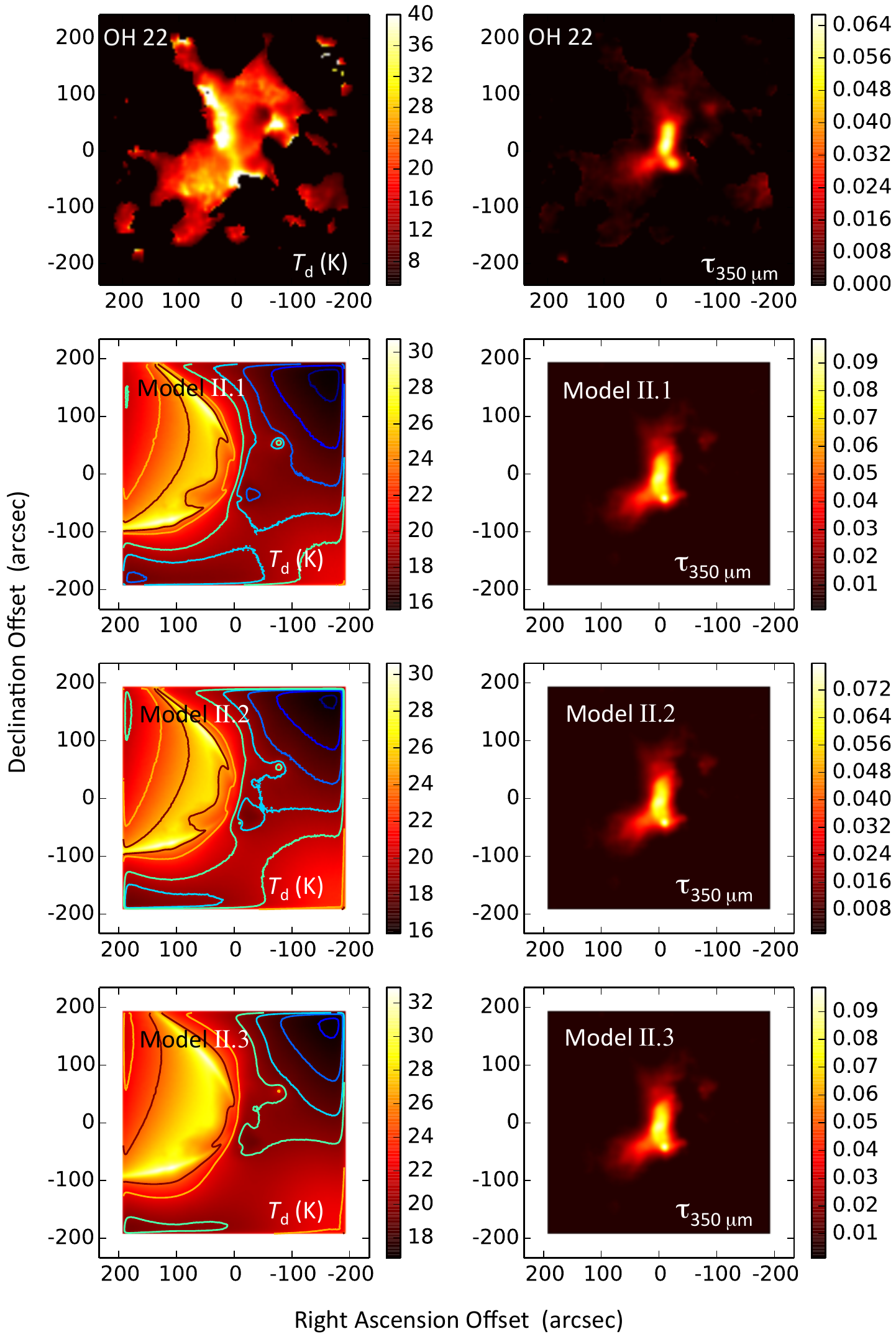}
        \caption{Distributions of dust temperature (left) and 350\,\um\ optical depth (right). {\bf Top:} 1D model using opacity curve No.\,22 of \citet[][Table\,B.1]{ossenkopf1994}, where $\beta=1.89$. {\bf Second from above:} 3D model (cf. Fig.\,\ref{TL_model}) accounting for gradients in the core and using a slightly modified curve No.\,22 with $\beta=1.99$ (model II.1). The contours correspond to 16, 17, 18, 19, 20, 25, and 30\,K. {\bf Second from below:} The same 3D model geometry, but with the opacity curve of \citet{zubko2004} that includes PAHs and for $\beta=2.1$ (model II.2, see the text).  {\bf Bottom:} Model\,II.3 has the same geometry as before, but uses opacities from \citet{weingartner2001} for $R_{\rm V}=5.5$ and PAHs (see Fig.\,\ref{TL_model}).}
  \label{dustT}
\end{figure*}


\subsection{The $g2d$ in \roa}

Table\,\ref{opacs} compares the results of different approaches to the dust modelling. Derived absolute masses can vary quite substantially, from 5 to 40\,\msun. However, opacity-weighted, normalized masses agree to within 10\% of each other, with an average value of 1.1 over an interval of 0.4 to 1.4.

To determine the absolute gas-to-dust mass ratio, $g2d$, is therefore not a trivial task. For example, if we choose the result of the non-equilibrium model\,2.III, the map average is only slightly sub-canonical,  $g2d=88$ (Fig.\,\ref{g2d}). This value would directly agree with that derived from X-ray work by \citet{vuong2003}, who found $g2d=95-80$ within the range $R_{\rm V}= 4-6.\,$\footnote{Other parameters include the metallicity $Z/Z_{\odot}=0.9 - 1.0$ and the minimum grain size $a_{\rm min}=0.025 - 0.07$\,\um.}

However, on scales of the individual sub-regions, this number would not be truly representative:  Over tens of seconds of arc, $g2d$ takes median values of $145^{+144}_{-105}$ around N\,2 in the north, of $14^{+28}_{-10} $ in the central regions of SM\,1, and of $104^{+97}_{-43}$ in the southeast, near N\,6. Taken literally, local $g2d$ values in \roa\ imply fractions of the dust mass from lower than 0.5\% to more than 10\%. In particular, the deviating, low values around the core SM\,1 were already remarked upon by \citet{bergman2011b}.

To bring the central $g2d$ into line with the widely accepted value of 100 would require the downward adjustment of the derived gas mass by a factor of ten. This is not a realistic option on the basis of the ALI modelling,
however, and must be discarded. Similarly, reducing the dust mass by an order of magnitude would result in $g2d$ of several thousand in the north and southeast, an option that is not convincing either. On the other hand, an uneven projected distribution of gas and dust may be a solution. In Fig.\,\ref{deconvol}, maps of gas and dust at 7\asec\ resolution are displayed, from which it becomes apparent that the continuum radiation peaks at positions offset from those of the gas. A low $g2d$ would naturally result.

\subsection{Distribution of gas and dust in \roa}

The C$^{18}$O\,(3-2)  line and the LABOCA continuum band are close in frequency, at about 850\,\um\ wavelength, and the maps of both have been observed with the same telescope (APEX), meaning
that these data share the same angular resolution of \about\,20\asec. Our SABOCA continuum map at 350\,\um\ has also been obtained with APEX, albeit at the higher resolution of 7\asec. Both APEX continuum maps share the same offset centre and internally agree excellently well (Sect.\,4.2). The nominal offset of the C$^{18}$O\,(3-2) map \citep{liseau2010} is at (\asecdot{$-45$}{$1$}, \asecdot{$+3$}{$0$}), which was applied to the line data to align with the continuum maps. As already mentioned, the pointing accuracy of the APEX telescope is 2\asec\ rms. 

We applied the program IMAGINE, which is based on an in-house developed algorithm \citep{rydbeck2008}, to the spatially oversampled C$^{18}$O\,(3-2) data. The number of noise channels in the line spectra, in each positional pixel, is large, which allows an accurate description of the statistics of the noise distribution. In addition, the spectra have a high S/N. These properties of the data are used by IMAGINE to deconvolve high-quality data, increasing the resolution by up to  a factor of  three. The deconvolved  C$^{18}$O\,(3-2) map of the integrated line intensity \tadv\ for the gas has then effectively the same resolution as the 350\,\um\ continuum data for the dust (Fig.\,\ref{deconvol}). It is gratifying to notice that the point source VLA\,1623, which is detected in both data sets, is at the same position in both maps. Furthermore, as a final sanity check, we degraded the high-resolution result with the APEX beam at 850\,\um, and this procedure did indeed recover the observed (input) map. We are therefore confident that any differences in the intensity distributions and on spatial scales exceeding $3\sigma\,(=6$\asec) of the pointing accuracy are real. 

This gas-dust dichotomy is not immediately apparent in the presented maps involving \ntwohp, since these have a much lower resolution, governed by the (J=6-5) observations with the roughly 40\asec-beam of {\it Herschel}. However, the (3-2) data have been obtained with APEX and for the same coordinates as the LA-/SABOCA maps. Using IMAGINE for the \ntwohp\,(3-2) map then yields a similar resolution as the SABOCA image and the sharpened C$^{18}$O\,(3-2) map. The result is shown in the right panel of Fig.\,\ref{deconvol}. The clear offset, by 12\asec, between the continuum source VLA\,1623 and an intensity peak in \ntwohp  is at first astonishing . Similar mismatches are also found elsewhere in these images, and they have also been noticed by others \citep{difrancesco2004,friesen2014}. We therefore conclude that the spatial non-coincidence between moleculer gas and dust is real. However, since the \ntwohp\ emission ridge is much narrower than that in C$^{18}$O, a depression in the distribution is not evidenced by the present image. This may require data  at higher resolution.

From Fig. \ref{deconvol}, it is immediately evident that the dust continuum and the molecular line emission have different spatial distributions in the sky. Maximum line emission in C$^{18}$O\,(3-2) is observed in a ridge closest to the stellar heating source S\,1, whereas the continuum is mostly pronounced offset to the west, where all dust models, both 1D and 3D, indicate dust temperatures $T_{\rm dust} \ge 20$\,K. In contrast, the spectral line analysis points toward gas temperatures that are clearly much lower than twenty Kelvin \citep[see also, e.g.,][]{stamatellos2007}. This
might effectively have created the ``hole'' of the C$^{18}$O emission, which might be due to depletion of the CO gas onto the dust.  At $T_{\rm kin}< 25\,K$, carbon monoxide is mostly  frozen out \citep{sandford1993a,oberg2005}, whereas diazenylium is possibly much less affected because it mainly remains in the gas phase. It is not clear, however, how \ntwohp\ would react
at temperatures  even below 15\,K \citep{oberg2005}. A selective freeze-out scenario would not explain the low $g2d$ that is based on \ntwohp\ observations. On the other hand, if all molecules are frozen out, there would be no contradiction.\footnote{\citet{sandford1993b} found evidence for frozen \molh\ in the \roc. However, this result was retracted in 2000, Sci., 287, 976.}
 
\begin{figure*}
 \resizebox{\hsize}{!}{
    \rotatebox{00}{\includegraphics{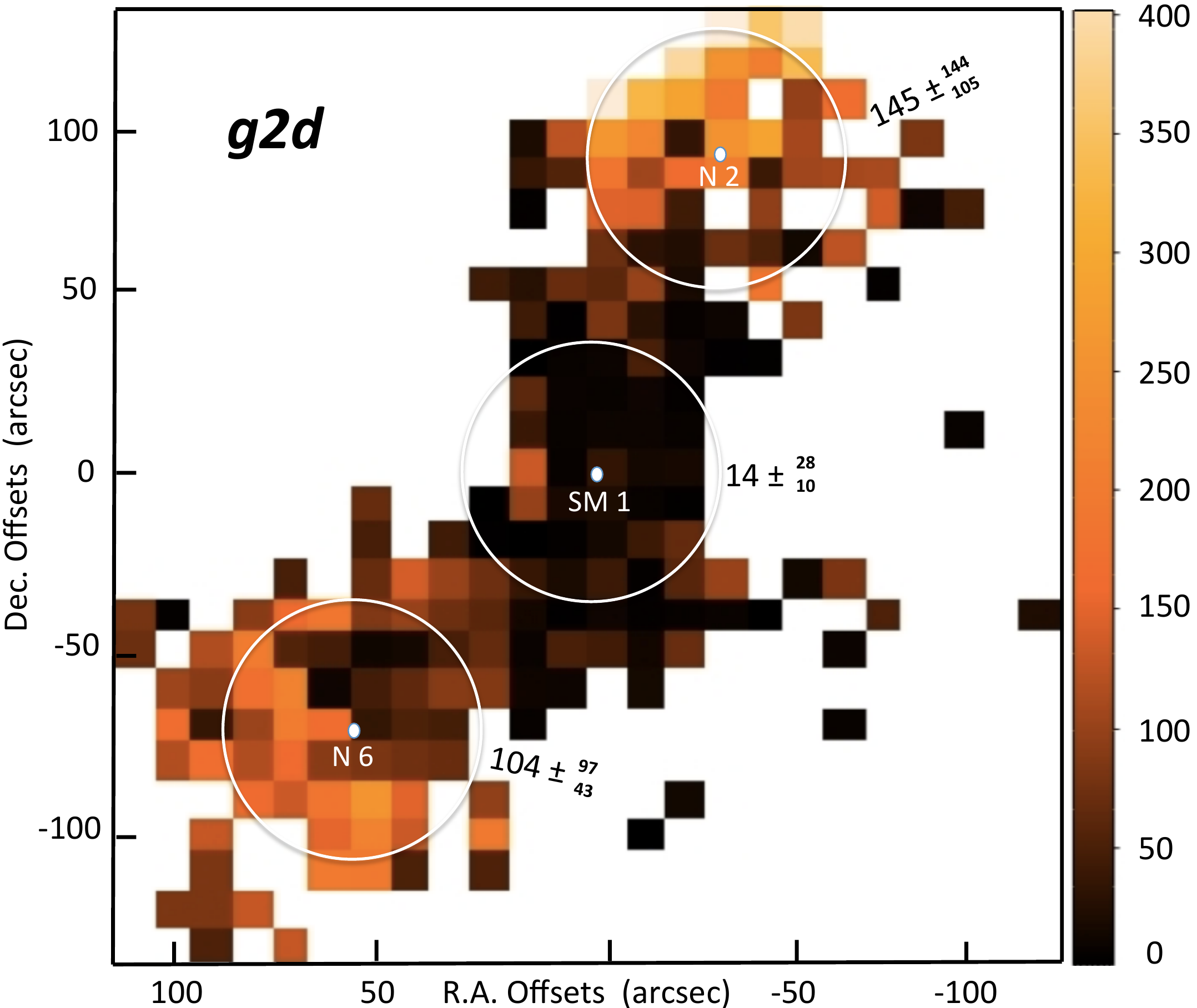}}    
                        }
  \caption{Distribution of the gas-to-dust ratio, $g2d$, in \roa. The gas mass estimate is based on the \ntwohp\ data and that of the dust on model\,II.3. The median values and their absolute deviations have been calculated inside the white circles around the sources in the north, N\,2, the centre, SM\,1, and the south, N\,6 \citep[see][]{difrancesco2004}. On the other hand, the average refers to all pixels of  the map, with the result  $g2d_{\rm ave}=88$.
       }
  \label{g2d}
\end{figure*}

\begin{figure*}
 \resizebox{\hsize}{!}{
    \rotatebox{00}{\includegraphics{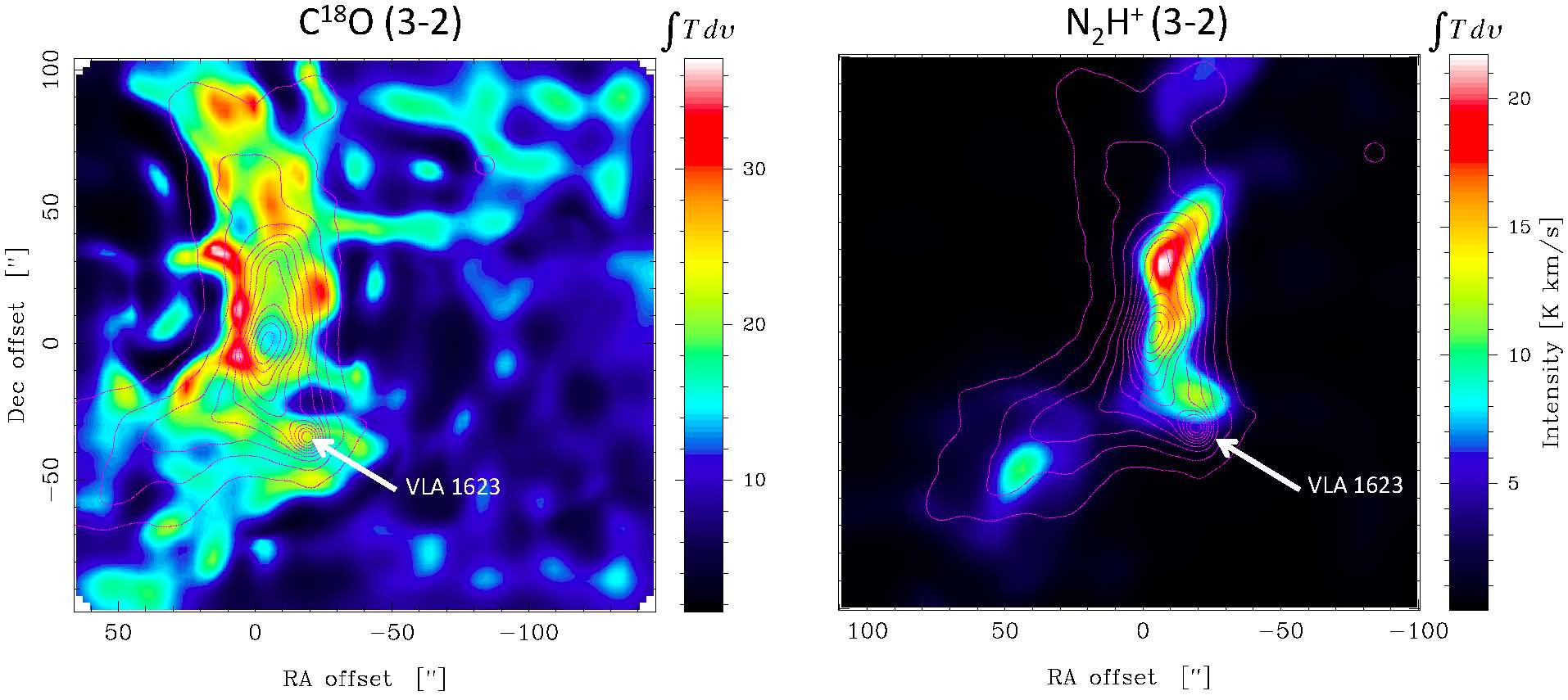}}
                        }
  \caption{{\bf Left:}  C$^{18}$O\,(3-2) map at 850\,\um, deconvolved to the same resolution (7\asec) as that of the SABOCA 350\,\um\ continuum data, shown as contours (from 2.5 to 30 in increments of 2.5\,Jy/beam). The Class\,0 source VLA\,1623 is clearly present in both the line and the continuum data. {\bf Right:}  Same as the left frame, but for \ntwohp\,(3-2) line and 350\,\um\ continuum radiation. The colour coding of the line intensity is given by the bar next to each image.
     }
  \label{deconvol}
\end{figure*}


\section{Conclusions}

In the FIR/submm, observed point sources exhibit a dependence of the dust parameter $\beta$ on evolutionary status (SED class), $\beta = \beta(t)$. Values decrease monotonically with age. For the pre-stellar core SM\,1, the continuum emission is entirely dominated  by the thick envelope and $\beta=2$, whereas in pre-main-sequence stellar objects of Class\,III, the envelope is lost and  the emission originates only from the circumstellar disk, $\beta \sim 0$. Objects of intermediate type display intermediate $\beta$-values. If not merely a matter of contrast, this may imply that grain growth primarily occurs in the protoplanetary disks, not in the natal cloud cores, since low values of $\beta$ are ascribed to large grains. 

\roa\ is exposed to uneven irradiation by stellar sources, heating the core primarily from the outside. We examined different dust emission models, including both 1D and 3D geometries, without and with thermal balance and with non-equilibrium conditions. Mass estimates span a factor of seven, but do agree to within a factor of two when proper weighting by the absolute mass opacity $\kappa_{\nu}$ is applied.

To model the observed gas emission in \ntwohp\ lines, best-fit solutions were obtained by combing a multi-dimensional parameter space. These provided excellent fits to the spatial distribution of the line intensity and the line shape. Combining the results for the gas with those for the dust yielded a global gas-to-dust mass ratio that is only slightly sub-canonical, $g2d=88$. However, on local scales, this hardly applies anywhere, with ratios much higher than \powten{2} in the north and south and significantly lower than that in the central regions. There, gas and dust appear
to be spatially segregated. At high angular resolution, a hole  of C$^{18}$O\,(3-2) emission is accompanied by maximum emission in the dust continuum. At lower resolution, this may result in a misleading $g2d$, and perhaps indicate that \ntwohp\ is frozen out as well. 

%

\begin{acknowledgement} We thank the anonymous referee for the very thoughtful report that led to a number of clarifications in the manuscript. Fran\c cois Lique kindly mad available to us the collision rate data for \ntwohp\ prior to publication, which we hereby gratefully acknowledge. R.L. enjoyed interesting discussions with John H. Black. The help by the APEX staff with our ground based observations is very much appreciated. We thank the Swedish National Space Board (SNSB)  for the continued support of  our {\it Herschel}-projects. The computations were performed on resources at Chalmers Centre for Computational Science and Engineering (C3SE) provided by the Swedish National Infrastructure for Computing (SNIC). 
\end{acknowledgement}

\bibliographystyle{aa}
\bibliography{refs}

\appendix
\section{Estimation of dust masses in homogenous and isothermal media}

The modification we made to the commonly adopted procedure is based on the fact that we do not a priori assume only a single dust temperature and grain opacity. Rather, these quantities are parameters to be fit iteratively by comparison with submillimeter observations. The value of the optical depth at any frequency is  obtained and tested a posteriori. The specific intensity $I_{\nu}$, in \ecshzsr, along the line of sight of optically thin thermal emission from dust particles can be written as

\begin{equation}
I_{\nu} = \int\!\kappa_{\nu}\,n_{\rm d}\,m_{\rm d}\,B_{\nu}(T_{\rm d})\,ds,\hspace{1cm}\tau_{\nu} \ll 1
\label{intensity}
,\end{equation}

\noindent
where $\kappa_{\nu}$ is the dust mass opacity, in cm$^2$\,g$^{-1}$, at the frequency $\nu$, in Hz, $n_{\rm d}$ is the volume density of dust particles, in \cmthree, $m_{\rm d}$ is the mass of a dust grain, in g, and $B_{\nu}(T_{\rm d})$  is the Planck function,  in \ecshzsr, at $\nu$ and  for the dust temperature $T_{\rm d}$, in Kelvin. 

For constant $\kappa_{\nu}$, $m_{\rm d}$ and $T_{\rm d}$ along a ray, the integral is only over the volume density $n_{\rm d}$, meaning that it is the column density of dust grains, $\int\!n_{\rm d}\,ds=N_{\rm d}$. The surface density of dust particles is $\Sigma_{\rm d}=m_{\rm d}\,N_{\rm d}$. Identification in Eq.\,\ref{intensity} yields

 \begin{equation}
\Sigma_{\rm d} = \frac {I_{\nu}}{\kappa_{\nu}\,B_{\nu}(T_{\rm d})}.
\label{surfden}
\end{equation}

From Eq.\,\ref{surfden} it is  apparent that the optical depth through the dust is given by 

\begin{equation}
\tau_{\nu} = \kappa_{\nu}\,\Sigma_{\rm d}.
\label{tau}
\end{equation}

The assumption that $\tau_{\nu} \ll 1$ will have to be justified a posteriori.  Applying Eq.\,\ref{surfden} to observations $I_{\nu 1}$ at a given frequency $\nu_1$, together with $\kappa_{\nu 1}$ from a particular opacity curve ($j=1,\,2,\,3,\,....$), yields a multitude ($i=1,\,2,\,3,\,...$) of solutions for both $\Sigma_{\rm d}$ and $T_{\rm d}$, that is,
\begin{equation}
\Sigma_{\rm d}(i, j) = \frac {I_{\nu 1}}{\kappa_{\nu 1}(j)\,B_{\nu 1}[T_{\rm d}(i, j)]}.
\label{mass}
\end{equation}

This multitude is limited by the range of examined temperatures, viz.  $T_{\rm d} = 4$ to 2000\,K, where the first value refers to the average equilibrium temperature of  grains irradiated by the interstellar radiation field in the solar neighbourhood\footnote{For \roa, the (anisotropic) radiation field is more intense by two orders of magnitude \citep{liseau1999}.} and the second to approximate dust evaporation temperatures.  Similarly, the number of discretized temperature values, $T_{\rm d}(i,\,j)$, that need to be computed is ultimately limited by the measuring accuracy of $I_{\nu}$. 

This multitude can be further reduced, in fact to unity, since, for any given opacity curve, only one value of $T_{\rm d}$ (and $\Sigma_{\rm d}$) will simultaneously satisfy Eq.\,\ref{surfden} for data at another frequency, for example, $I_{\nu 2}$, viz.

\begin{equation}
\Delta I_{\nu 2}(j) = I_{\nu 2}^{\rm obs} -  \Sigma_{\rm d}(i, j)\,\kappa_{\nu 2}(j)\,B_{\nu 2}[T_{\rm d}(i,j)],
\label{temp}
\end{equation}

\noindent
where we let $\Delta I_{\nu 2}(j) \rightarrow 0$. Eq.\,\ref{temp} yields for each opacity curve, designated by $j$, a unique solution for $T_{\rm d}(j)$ and $\Sigma_{\rm d}(j)$. It only remains to select the correct $\kappa$-curve. Minimization of the data for a third frequency, $I_{\nu 3}$, could in principle be used to achieve this. Hence

\begin{equation}
\Delta \kappa_{\nu 3}(j) = \kappa_{\nu 3}(j) - \frac {I_{\nu 3}} {\Sigma_{\rm d}(j)\,B_{\nu 3}[T_{\rm d}(j)]}.
\label{kappa}
\end{equation}

Minimizing $\Delta \kappa_{\nu}$ then provides the unique solution for the temperature $T_{\rm d}$, the surface density $\Sigma_{\rm d}$ and the opacity law of the dust, $\kappa=\kappa({\nu})$. Finally, the assumption of low optical depth at the different frequencies of this solution is then verified a posteriori (Eq.\,\ref{tau}). 

The column density of the gas, $N_{\rm g}=N$(H), needs to be determined from other observations. This, finally, leads to the gas-to-dust mass ratio from

\begin{equation}
g2d = \frac {\mu\,m_{\rm H}\,N_{\rm g}}{\Sigma_{\rm d}}
\label{mgd}
,\end{equation}

\noindent
where $\mu=2.4$ is the mean molecular weight. The derived dust opacity curve has been computed for a particular volume density of the gas, $n^{\rm d}$(H) and this ought to be consistent with that determined from the spectral line fitting,  $n^{\rm g}$(H). The spatial averaging is done over each pixel, so that the total dust mass, $M_{\rm d}$, is then obtained from the summation of the surface density per pixel over all pixels.


\end{document}